\documentclass[11pt]{article}

\usepackage[round]{natbib}
\usepackage{amssymb}
\usepackage{amsmath}
\usepackage{amsthm}
\usepackage{graphicx}
\usepackage[margin=1in]{geometry}
\usepackage{color}
\usepackage{authblk}
\newcommand{\bzero}{\mathbf{0}}

\newcommand{\bt}{\mathbf{t}}
\newcommand{\btnoti}{\mathbf{t}_{-i}}
\newcommand{\bT}{\mathbf{T}}
\newcommand{\btni}{\mathbf{t}_{\mathcal{N}_i}}
\newcommand{\btnj}{\mathbf{t}_{\mathcal{N}_j}}

\newcommand{\bTnj}{\mathbf{T}_{\mathcal{N}_j}}
\newcommand{\bw}{\mathbf{w}}
\newcommand{\G}{\mathcal{G}}

\newcommand{\Cmax}{C_{max}}
\newcommand{\Ni}{\mathcal{N}_i}
\newcommand{\N}{\mathcal{N}}

\newcommand{\Gni}{G_{\mathcal{N}_i}}
\newcommand{\Gnj}{G_{\mathcal{N}_j}}

\newcommand{\gni}{g_{\mathcal{N}_i}}
\newcommand{\Gnit}{G_{\mathcal{N}_i}^{\bt}}
\newcommand{\GnT}{G_{\mathcal{N}}^{\bT}}
\newcommand{\GniT}{G_{\mathcal{N}_i}^{\bT}}
\newcommand{\gnit}{g_{\mathcal{N}_i}^{\bt}}
\newcommand{\hnit}{h_{\mathcal{N}_i}^{\bt}}
\newcommand{\gnjt}{g_{\mathcal{N}_j}^{\bt}}
\newcommand{\Gnjt}{G_{\mathcal{N}_j}^{\bt}}
\newcommand{\GnjT}{G_{\mathcal{N}_j}^{\bT}}

\newcommand{\GnkT}{G_{\mathcal{N}_k}^{\bT}}

\newcommand{\gnjtj}{g_{\mathcal{N}_j}^{\bt_j}}

\newcommand{\GnjTj}{G_{\mathcal{N}_j}^{\bT_j}}

\newcommand{\GnkTk}{G_{\mathcal{N}_k}^{\bT_k}}

\newcommand{\hniti}{h_{\mathcal{N}_i}^{\bt_i}}
\newcommand{\cT}{\mathcal{T}}
\newcommand{\MG}{\mathtt{MG}}
\newcommand{\PE}{\mathtt{PE}}
\newcommand{\PEhat}{\widehat{\PE}}
\newcommand{\F}{\mathcal{F}}
\newcommand{\E}{\mathbb{E}}
\newcommand{\indep}{\mathrel{\text{\scalebox{1.07}{$\perp\mkern-10mu\perp$}}}}
\DeclareMathOperator*{\argmin}{arg\,min}
\DeclareMathOperator*{\argmax}{arg\,max}

\newcommand{\ind}{\mathbb{I}}
\newcommand{\com}[1]{&&\mbox{(#1)}}
\newcommand{\dist}[2]{\ind[S(#1) \neq S(#2)]}
\newcommand{\wdist}[2]{\bw^T\dist{#1}{#2}}

\newtheorem{proposition}{Proposition}

\title{Almost-Matching-Exactly for Treatment Effect Estimation under Network Interference}

\author[1]{M. Usaid Awan}
\author[2]{Marco Morucci} 
\author[3]{Vittorio Orlandi}
\author[4]{Sudeepa Roy} 
\author[3,4,5]{Cynthia Rudin} 
\author[3]{Alexander Volfovsky}
\affil[1]{Department of Economics, Duke University}
\affil[2]{Department of Political Science, Duke University}
\affil[3]{Department of Statistical Science, Duke University}
\affil[4]{Department of Computer Science, Duke University}
\affil[5]{Department of Electrical and Computer Engineering, Duke University}
\date{}

\begin{document}
\maketitle
\begin{abstract}
We propose a matching method that recovers direct treatment effects from randomized experiments where units are connected in an observed network, and units that share edges can potentially influence each others' outcomes. Traditional treatment effect estimators for randomized experiments are biased and error prone in this setting. Our method matches units almost exactly on counts of unique subgraphs within their neighborhood graphs. The matches that we construct are interpretable and high-quality. Our method can be extended easily to accommodate additional unit-level covariate information. We show empirically that our method performs better than other existing methodologies for this problem, while producing meaningful, interpretable results.
\end{abstract}
%

%




\section{INTRODUCTION}

Randomized experiments are considered to be the gold standard for estimating causal effects of a treatment on an outcome. Typically, in these experiments, the outcome of a unit is assumed to be only affected by the unit's own treatment status, and not by the treatment assignment of other units \citep{cox1958planning,rubin1980}. However, in many applications -- such as measuring effectiveness of an advertisement campaign or a teacher training program -- units interact, and ignoring these interactions results in poor causal estimates \citep{halloran1995, sobel2006}. We propose a method that leverages the observed network structure of interactions between units to account for treatment interference among them. 

We study a setting in which a treatment has been uniformly randomized over a set of units connected in a network, and where treatments of connected units can influence each others' outcomes. The development of methods for this setting is a relatively new field in causal inference methodology, and only few approaches for it have been proposed \citep[e.g.,][]{van2014, aronow2017, sussman2017}. 

In this paper,
we propose a method that leverages matching \citep{rosenbaum1983central} to recover direct treatment effects from experiments with interference. Our method makes several key contributions to the study of this setting:
First, {\it our method explicitly leverages information about the network structure of the experimental sample to adjust for possible interference while estimating direct treatment effects}. 
Second, unlike other methods, matching allows us to \textit{nonparametrically} estimate treatment effects, without the need to specify parametric models for interference or outcomes. Third, matching produces highly interpretable results, informing analysts as to which features of the input data were used to produce estimates. More specifically, {\it we match on features of graphs that are easy to interpret and visualize}. 

In our setting, units experience interference according to their neighborhood graphs -- the graphs defined by the units they are directly connected to. Units with similar neighborhood graphs will experience similar interference. For example, the educational outcome of a student randomly assigned to an extra class depends on whether or not her friends are also assigned to that class, and not just on how many: the specific structure of the student's friendship circle will influence whether or not study groups are formed, how information is shared, how much attention the treated student will devote to the class, and so on. All of this will impact the overall educational outcomes of interest. 

Because of this, matching units with similar neighborhood graphs together will enable us to recover direct treatment effects even under interference. We match units' neighborhood graphs on counts of subgraphs within them, as graphs with similar counts of the same unique subgraphs are naturally likely to be similar. From there, we construct matches on individuals with similar sets of important subgraphs; here, the set of important subgraphs is learned from a training set.
We generalize the Almost-Matching-Exactly (AME) framework \citep{DAME, FLAME} to match units on subgraphs in experimental settings. We do this by constructing graph-based features that can explain both the interference pattern in the experiment and predict the underlying social network. We demonstrate that our method performs better than other methods for the same problem in many settings, while generating interpretable matches.

The paper will proceed as follows: In Section \ref{Sec:Methodology}, we make explicit the assumptions underpinning our framework, and outline our matching approach to estimating direct treatment effects. In Sections \ref{Sec:Experiments} and \ref{Sec:Application}, we evaluate the effectiveness of our method on simulated and real-world data. Theoretical evaluation of our approach is available in the appendix.


\subsection{Related Work}
Work on estimating causal effects under interference between units has three broad themes.
First, there has been a growing body of work on the design of novel randomization schemes to perform causal inference under interference \citep{liu2014large,sinclair2012detecting,duflo2003role,basse2018model}. Some of this work makes explicit use of observed network structure to randomly assign treatment so as to reduce interference \citep{ugander2013, toulis2013, eckles2016, eckles2017, jagadeesan2017designs}. These methodologies are inapplicable to our setting as they require non-uniform treatment assignment, whereas in our setting we wish to correct for interference after randomization.
Second, there is work on estimating direct treatment effects in experiments under interference, and after conventional treatment randomization, similar to our setting. Some existing work aims to characterize the behavior of existing estimators under interference \citep{manski2013, savje2017average}. Other approaches lay out methods based on randomization inference to test a variety of hypotheses under interference and treatment randomization \citep{rosenbaum2007, aronow2012general, athey2018}. Some of these approaches mix randomization inference and outcome models \citep{bowers2013}. For the explicit problem of recovery of treatment effects under interference, \cite{aronow2017} provide a general framework to translate different assumptions about interference into inverse-probability estimators, and \cite{sussman2017} give linearly unbiased, minimum integrated-variance estimators under a series of assumptions about interference. These methods either ignore explicit network structure, or require probabilities under multiple complex sampling designs to be estimated explicitly.  
Finally, there have been studies of observational inference under network interference \citep{van2014, liu2016, ogburn2017vaccines, forastiere2016}. However, recovering causal estimates using observational data when units are expected to influence each other requires a structural model of both the nature of interference and contagion among units. 
\vspace{-2mm}
\section{METHODOLOGY}\label{Sec:Methodology}
We discuss our problem  and approach in this section.
\subsection{Problem Statement}
We have a set of $n$ experimental units indexed by $i$. These units are connected in a known graph $G = (V, E)$, where $V(G) = \{1, \dots, n\}$ is the set of vertices of $G$, and $E(G)$ is the set of edges of $G$. 
We disallow self-loops in our graph. 
We say that $H$ is a subgraph of $G$ if $V(H) \subseteq V(G)$ and $E(H) \subseteq E(G)$.  
Let $t_i \in \{0, 1\}$ represent the treatment indicator for unit $i$, $\bt$ represent the vector of treatment indicators for the entire sample, and $\bt_{-i}$ represent the treatment indicators for all units except $i$. Given a treatment vector $\bt$ on the entire sample (i.e., for all vertices in $G$),  we use $G^\bt$ to denote the {\em labeled graph}, where each vertex $i \in V(G)$ has been labeled with its treatment indicator $t_i$. In addition, we use $G_{P}$ to denote a graph induced by the set of vertices $P \subseteq V(G)$ on $G$, such that $V(G_P) = P$ and $E(G_P) = \{(e_1, e_2) \in E(G):\; e_1 \in P, e_2 \in P\}$. We use the notation $\Ni = \{j: (i, j) \in E(G)\}$ to represent the neighborhood of vertex $i$. The labeled neighborhood graph of a unit $i$, $\Gnit$, is defined as the graph induced by the neighbors of $i$, and labeled according to $\bt$. We also define $\btni$ to be the vector of treatment indicators corresponding to unit $i$'s neighborhood graph. A unit's response to the treatment is represented by its random potential outcomes $Y_i(\bt)$ = $Y_i(t_i, \bt_{-i})$. Unlike other commonly studied causal inference settings, unit $i$'s potential outcomes are now a function of both the treatment assigned to $i$, and of all other units' treatments. Observed treatments for unit $i$ and the whole sample are represented by the random variables $T_i$ and $\bT$ respectively. We assume that the number of treated units is always $n^{(1)}$, i.e., $\sum_{i=1}^n T_i = n^{(1)}$.

\textbf{A0: Ignorability of Treatment Assignment.} We make the canonical assumption that treatments are administered independently of potential outcomes, that is: $Y_i(t_i, \btnoti)\indep \bT$, and $0 <\Pr(T_i = 1) < 1$ for all units. In practice, we assume that treatment is assigned uniformly at random to units, which is possible only in experimental settings. As stated before, we \textbf{do not} make the canonical Stable Unit Treatment Value Assumption (SUTVA) \citep{rubin1980}, which, among other requirements, states that units are exclusively affected by the treatment assigned to them. We do not make this assumption because our units are connected in a network: it could be possible for treatments to spread along the edges of the network and to affect connected units' outcomes. We do maintain the assumption of comparable treatments across units, which is commonly included in SUTVA. 

Our causal quantity of interest will be the Average Direct Effect (ADE), which is defined as follows:
\begin{align}
    ADE &= \frac{1}{n} \sum_{i = 1}^n \E[Y_i(1, \bzero) - Y_i(0, \bzero)],
\end{align}
where $\bt_{-i} = \bzero$ represents the treatment assignment in which no unit other than $i$ is treated. The summand represents the treatment effect on unit $i$ when no other unit is treated, and, therefore, no interference occurs \citep{halloran1995}. 

\subsection{Framework}\label{Sec:Framework}
We outline the requirements of our framework for direct effect estimation under interference. We denote interference effects on a unit $i$ with the function $f_i(\bt):\{0, 1\}^n\mapsto\mathbb{R}$, a function that maps each possible treatment allocation for the $n$ units to the amount of interference on unit $i$. We will use several assumptions to restrict the domain of $f$ to a much smaller set 
(and overload the notation $f_i$ accordingly). To characterize $f$, we rely on the typology of interference assumptions introduced by \cite{sussman2017}. The first three assumptions (A0-A2) needed in our framework are common in the interference literature \citep[e.g.,][]{manski2013, toulis2013, eckles2016, athey2018}:

\textbf{A1: Additivity of Main Effects.} First, we assume that main treatment effects are additive, i.e., that there is no interaction between units' treatment indicators.  This allows us to write: 
\begin{align}
Y_i(t, \bt_{-i}) = t\tau_i + f_i(\btnoti) + \epsilon_i  \label{eq:additive}
\end{align}
where $\tau_i$ is the direct treatment effect on unit $i$, and $\epsilon_i$ is some baseline effect. \\
\textbf{A2: Neighborhood Interference.} We focus on a specific form of the interference function $f_i$ by assuming that the interference experienced by unit $i$ depends only on treatment of its neighbors. That is, if for two treatment allocations $\bt, \bt'$ we have
$\btni = \btni'$ then $f_i(\bt) = f_i(\bt')$. To make explicit this dependence on the neighborhood subgraph, we will write $f_i(\btni) \equiv 
f_i(\bt)$.\\
\textbf{A3: Isomorphic Graph Interference} We assume that, if two units $i$ and $j$ have \emph{isomorphic labeled neighborhood graphs}, then they receive the same amount of interference, 
denoting isomorphism by $\simeq$, $\Gnit \simeq \Gnjt \implies f_i(\btni) = f_j(\btnj) 
\equiv f(\Gnit) = f(\Gnjt)$. 
While Assumptions A1 and A2 are standard, 
A3 is new. This assumption allows us to study interference in a setting where units with similar neighborhood subgraphs experience similar amounts of interference. 

All our assumptions together induce a specific form for the potential outcomes, namely that they depend on neighborhood structure $\Gnit$, but not exactly who the neighbors are (information contained in $\Ni$) nor treatment assignments for those outside the neighborhood (information contained in $\btni$). Namely: 
\begin{proposition}{}\label{Thm:Parametric}
Under assumptions A0-A3, potential outcomes in (\ref{eq:additive}) for all units $i$ can be written as:
\begin{align}
    Y_i(t, \btnoti) &= t\tau_i + f(\Gnit) + \epsilon_i, \label{Eq:OutcomeModel}
\end{align}
where $\tau_i$ is the direct treatment effect on unit $i$, 
and $\epsilon_i$ is some baseline response. 

In addition, suppose that baseline responses for all units are equal to each other in expectation, 
i.e., for all $i$, $\E[\epsilon_i] = \alpha$. Then under assumptions A0-A3, for neighborhood graph structures 
$g_i$ of unit $i$ 
and treatment vectors $\bt$, the ADE is identified as:
\begin{align*}
ADE = \frac{1}{n^{(1)}}\sum_{i=1}^n \E\bigl[T_i \times &\bigl(\E[Y_i|\GniT \simeq g_i^\bt, T_i = 1]\\
&- \E[Y_i|\GnT \simeq g_i^\bt, T_i = 0]\bigr)\bigr],
\end{align*}
where $\GniT$ is the neighborhood graph of $i$ labelled according to the treatment assignment $\bT$.
\end{proposition}
The proposition (whose proof is in the appendix) states that the interference received by a unit is a function of each unit's neighborhood graph. Further, the outcomes can be decomposed additively into this function and the direct treatment effect on $i$. The proposition implies that the ADE is identified by matching each treated unit to one or more control units with an isomorphic neighborhood graph, and computing the direct effect on the treated using these matches. This effect is, in expectation over individual treatment assignments, equal to the ADE.  

\subsection{Subgraph Selection via Almost-Matching-Exactly}\label{Sec:AME}
Given Proposition 1 and the framework established in the previous section, we would ideally like to match treated and control units that have isomorphic neighborhood graphs. This would allow us to better estimate the ADE without suffering interference bias: for a treated unit $i$, if a control unit $j$ can be found such that $\Gnit \simeq \Gnjt$, then $j$'s outcome will be identical in expectation to $i$'s counterfactual outcome and can be used as a proxy. Unfortunately, the number of non-isomorphic (canonically unique) graphs with a given number of nodes and edges grows incredibly quickly \citep{harary1994} and finding such matches is infeasible for large graphs. We therefore resort to counting all subgraphs that appear in a unit's neighborhood graph and matching units based on the counts of those subgraphs. However, instead of \textit{exactly} matching on the counts of those subgraphs, we match treated and control units if they have \emph{similar} counts, since matching exactly on all subgraph counts implies isomorphic neighborhoods and is also infeasible. Further, absolutely exact matches may not exist in real networks. 

\begin{figure}[!htbp]
    \centering
    \includegraphics[width=0.7\linewidth]{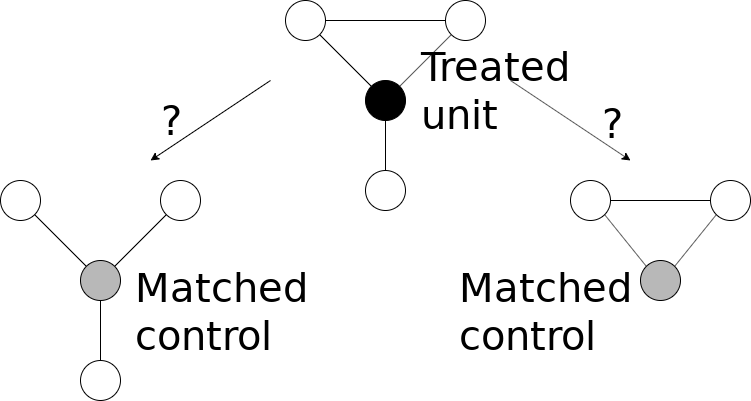}
    \caption{Inexact matching presupposes an ordering of feature importance; should the the treated ego (black) be matched to a control whose neighborhood graph 
    has the same number of units (left), or same number of triangles (right)?}
    \label{Fig:Subgraphs}
    \vspace{-2mm}
\end{figure}

Constructing inexact matches, in turn, requires a measure of relative graph importance. In Figure \ref{Fig:Subgraphs}, for example, there are two control units that the treated unit may be matched to; if triangles contribute more to the interference function, it should be matched to the right; otherwise, if degree and/or two-stars are more important, it should be matched to the left. Of course, these relative importance measures might depend on the problem and we would like to learn them. \\

It might be tempting to match directly on $f$, as that would lead to unbiased inference. However, we abstain from doing so for two reasons. Firstly, in practice, the true interference is unknown and we could only match on estimated values of $f$; this suffers from all the problems that afflict matching on estimated propensity scores without appropriate adjustments \citep{abadie2016} or parametric approximations \citep{rubin1996}. Such corrections or approximations do not currently exist for estimated interference functions and their development is an active area of research. Secondly, interpretability is a key component of our framework that would be lost matching on $f$-values; these values are scalar summaries of interference that depends on entire graphs. Estimating $f$ well would also likely require complex and uninterpretable nonparametric methods. In Section \ref{sec:trueinterference} of the appendix, we empirically compare matching units on $f$-values to our subgraph matching method via simulation. The loss of interpretability associated with matching on $f$ does not yield substantial gains in performance, even when using \emph{true} values of $f$ for matching, which is impossible in practice. 

Almost-Matching-Exactly (AME) \citep{FLAME, DAME, FLAMEIV} provides a framework for the above problem that is explicitly geared towards building interpretable, high-quality matches on discrete covariates, which in our setting are the counts of the treated subgraphs in the neighborhood. AME performs inexact matching while \textit{learning} importance weights for each covariate from a training set, prioritizing matches on more important covariates. In this way, it neatly addresses the challenge of inexact matching by learning a metric specific to discrete covariates (namely, a weighted Hamming distance).
Formally, AME matches units so as to optimize a flexible measure of match quality. For each treated unit $i$, solving the AME problem is equivalent to finding:
\begin{align}
\boldsymbol{\theta}^{i^*} &\in \argmax_{\boldsymbol{\theta} \in \{0, 1\}^p} \boldsymbol{\theta}^T\mathbf{w} \label{Eq:AME}\\
&\text{such that } \exists j: t_j = 0 \text{ and } \mathbf{x}_j \circ \boldsymbol{\theta} = \mathbf{x}_i \circ \boldsymbol{\theta} \nonumber
\end{align}
where $\circ$ denotes the Hadamard product, $\mathbf{w}$ is a vector of weights and $\boldsymbol{x}_i, \boldsymbol{x}_j$ are vectors of binary covariates for units $i$ and $j$ that we might like to match on. In our network interference setting, these are vectors of subgraph counts. The vector $\bw$ denotes the importance of each subgraph in causing interference. We will leverage both information on outcomes and networks to construct an estimate for it. 

We start by enumerating (up to isomorphism) all the $p$ subgraphs $g_1, \dots, g_p$ that appear in any of the $\Gnit, i \in 1, \dots, n$. The covariates for unit $i$ are then given by $S(\Gnit) = (S_1(\Gnit), \dots, S_p(\Gnit))$ where $S_k(\Gnit)$ denotes the number of times subgraph $g_k$ appears in the  subgraphs of $\Gnit$. These counts are then converted into binary indicators that are one if the count of subgraph $g_k$ in each unit's neighborhood is exactly $x$, for all $x$ observed in the data. Thus, units will be matched exactly if they have identical subgraph counts. 
We then approximately solve the problem in Equation \eqref{Eq:AME} to find the optimally important  set of subgraphs upon which to exactly match each treated unit, such that there is at least one control unit that matches exactly with the treated unit on the chosen subgraph counts. 
The key idea behind this approach is that we want to match units exactly on subgraph counts that contribute significantly to the interference function, trading off  exactly-matching on these important subgraphs with potential mismatches on subgraphs that contribute less to interference.

In practice, our implementation enumerates all subgraphs in each unit's neighborhood and stores the count of each pattern -- this is computationally challenging. There is a growing body of work on efficient counting algorithms for pre-specified small patterns (up to 4-5 nodes) but there is little research on fast methods to both enumerate and count all motifs in a graph \citep[e.g.,][]{pinar2017, marcus2010, hu2013}. Empirically, we see that this enumeration takes less than 30 seconds for 50 units. 

\paragraph{The FLAME Algorithm for AME.}
The Fast Large Almost Matching Exactly (FLAME) algorithm \citep{FLAME} approximates the solution to the AME problem. The procedure starts by exactly matching all possible units on all covariates. It then drops one covariate at a time, choosing the drop maximizing the match quality $\mathtt{MQ}$ at that iteration, defined:
\begin{align}
    \mathtt{MQ} = C \cdot \mathtt{BF} - \widehat{\mathtt{PE}}_Y.
    \label{Eq:MQ}
\end{align}
The match quality is the sum of a balancing factor $\mathtt{BF}$ and a predictive error $\widehat{\mathtt{PE}}_Y$, with relative weights determined by the hyper-parameter $C$. The balancing factor is defined as the proportion of treated units plus the proportion of control units matched at that iteration.
Introducing the balancing factor into the objective  has the advantage of encouraging more units to be matched, thereby minimizing variance of estimators \citep[see][]{FLAME}.   
In our setting, the second component of the match quality, predictive error, takes the form:
\begin{align}
    \widehat{\mathtt{PE}}_Y = \argmin_{h \in \mathcal{F}_1}\sum_i^n(Y_i - h(S(\Gnit) \circ \boldsymbol{\theta}, T_i))^2\label{Eq:PE}
\end{align} for some class of functions $\mathcal{F}_1$. It is computed using a holdout training set and discourages dropping covariates that are useful for predicting the outcome. In this way, FLAME strikes a balance between matching many units and ensuring these matches are of high-quality. By using a holdout set to determine how useful a set of variables is for out-of-sample prediction, FLAME learns a measure of covariate importance via a weighted Hamming distance. Specifically, it learns a vector of importance weights $\mathbf{w}$ for the different subgraph counts that minimizes $\wdist{\Gnit}{\Gnjt}$, where $\mathbb{I}[S(\Gnit) \neq S(\Gnjt)]$ is a vector whose $k^{\textrm{th}}$ entry is 0 if the labeled neighborhood graphs of $i$ and $j$ have the same count of subgraph $k$, and 1 otherwise.

To this match quality term, we add a \textit{network fit} term to give subgraphs more weight that are highly predictive of overall network structure. We fit a logistic regression model in which the edges $(i, j)$ between units $i, j$ are independent given $\mathcal{N}_i, \mathcal{N}_j$, and dependent on the subgraph counts of units $i$ and $j$: \[(i, j) \stackrel{iid}{\sim} \textrm{Bern}(\textrm{logit}(\beta_1^TS(\Gnit) + \beta_2^TS(\Gnjt)))\]
To the match quality in the original formulation, we then add $\widehat{\mathtt{PE}}_G$, defined to be the \emph{AIC} \citep{AIC74} of this fitted model, weighted by a hyperparameter $D$. Therefore, at each iteration:
\begin{align*}
    \mathtt{MQ} = C \cdot \mathtt{BF} - \widehat{\mathtt{PE}}_Y + D \cdot \widehat{\mathtt{PE}}_G.
\end{align*}
Thus, we penalize not only subgraph drops that impede predictive performance or making matches, but also those that make the observed network unlikely.
$\PEhat_G$ represents the empirical prediction error of the chosen set of statistics for the observed graph: if $\PEhat_G$ is low, then the chosen subgraphs do a good job of predicting the observed graph.
This error is also evaluated at a minimum over another class of prediction functions, $\F_2$. This term in the AME objective is justified by Assumption A3: units that have isomorphic labeled neighborhood graphs should experience the same amount of interference, and subgraph counts should be predictive of neighborhood graph structure. 

Our approach to estimating the ADE is therefore as follows. (1) For each unit $i$, count and label all types of subgraphs in $\Gnit$. (2) Run FLAME, encouraging large numbers of matches on subgraph counts, while using the covariates that are most important for predicting the outcome and the network. (3) Estimate ADE as $\widehat{ADE}$, by computing the difference in means for each matched group and then averaging across matched groups, weighted by their size. Since our approach is based on FLAME, we call it \emph{FLAME-Networks}.

\textbf{Extensions.} FLAME-Networks immediately extends to handling unit-level covariate information for baseline adjustments; we simply concatenate subgraph information and covariate information in the same dataset and then make almost-exact matches. FLAME-Networks will automatically learn the weights of both subgraphs and baseline covariates to make  matches that take both into account. 
Another straightforward extension considers interference not just in the immediate neighborhood of each unit, but up to an arbitrary number of hops away. To extend FLAME-Networks in this way, it is sufficient to enumerate subgraphs in the induced neighborhood graph of each unit $i$  where the vertices considered are those with a path to $i$ that is at most $k$ steps long. Given these counts, our method proceeds exactly as before. 

\section{EXPERIMENTS}\label{Sec:Experiments}

We begin by evaluating the performance of our estimator in a variety of simulated settings, in which we vary the form of the interference function. We find that our approach performs well in many distinct settings. In Section \ref{sec:matchquality} of the appendix, we also assess the quality of the matches constructed.

We simulate graphs from an Erd\H{o}s-R\`enyi model: $G \sim \textrm{Erd\H{o}s-R\`enyi}(n, q)$, by which every possible edge between the $n$ units is created independently, with probability $q$. 
In Sections \ref{sec:graphcluster} and \ref{sec:realnetwork} of the appendix, we also perform experiments on cluster-randomized and  real-world networks. Treatments for the whole sample are generated with $\Pr(\bT = \bt) = \binom{n}{n^{(1)}}^{-1}$, where $n^{(1)}$ is the number of treated units.  Outcomes are generated according to $Y_i(t, \bt_{-i})  =  \mathbf{t}\tau_i + f(\Gnit) + \epsilon_i$, where $\epsilon \sim N(\mathbf{0}, I_n)$ represents a baseline outcome; $\tau_i \sim N(\mathbf{5}, I_n)$ represents a direct treatment effect, and $f$ is the interference function. In Section \ref{sec:heteronetwork} of the appendix, we consider a setting in which the errors are heteroscedastic. For the interference function, we use additive combinations of the subgraph components in Table \ref{Tab:Subgraphs} and define $m_{ip}, p = 1, \dots, 7$ to be the counts of feature $p$ in $G_{\Ni \cup \{i\}}^\bt$. Lastly, the counts of each component are normalized to have mean 0 and standard deviation 1. 
\begin{table}[] 
\centering
\begin{tabular}{rp{7cm}}
\hline
$d_i$: & The treated degree of unit $i$\\
$\Delta_i$: & The number of triangles in $G_{\Ni \cup \{i\}}^\bt$ with at least one treated unit.\\
$\bigstar_i^k$: & The number of $k$-stars in $G_{\Ni \cup \{i\}}^\bt$ with at least one treated unit. \\
$\dagger_i^k$: & The number of units in $G_{\Ni \cup \{i\}}^\bt$ with degree $\geq k$ and at least one treated unit among their neighbors.\\
$B_i$: & The vertex betweenness of unit $i$.\\
$C_i$: & The closeness centrality of unit $i$.\\
\hline
\end{tabular}
\caption{Interference components used in experiments; see the appendix for more details.}
\vspace{-3mm}
\label{Tab:Subgraphs}

\end{table}
We compare our approach with different methods to estimate the ADE under interference: 

\textbf{Na\"ive}.  The simple difference in means between treatment and control groups assuming no interference.  \\
\textbf{All Eigenvectors}. Eigenvectors for the entire adjacency matrix are computed with every treated unit matched to the control unit minimizing the Mahalanobis distance between the eigenvectors, weighing the $k$'th eigenvector by $1/k$. The idea behind this estimator is that the eigendecomposition of the adjacency matrix encodes important information about the network and how interference might spread within it. \\
\textbf{First Eigenvector}. Same as \textbf{All Eigenvectors} except units are matched only on their values of the largest-eigenvalue eigenvector. \\
\textbf{Stratified Na\"ive}. The stratified na\"ive estimator as discussed by \citet{sussman2017}. A weighted difference-in-means estimator where units are divided into strata defined by their treated degree (number of treated vertices they are connected to), and assigned weight equal to the number of units within the stratum in the final difference of weighted averages between treated and control groups. \\
\textbf{SANIA MIVLUE}. The minimum integrated variance, linear unbiased estimator under assumptions of symmetrically received interference and additivity of main effects, when the priors on the baseline outcome and direct treatment effect have no correlation between units; proposed by \citet{sussman2017}.\\
\textbf{FLAME-Networks}. Our proposed method. 
In all simulations, the two components of the PE function are weighted equally, and a ridge regression is used to compute outcome prediction error.
\vspace{-2mm}
\subsection{Experiment 1: Additive Interference}
\begin{figure}[htbp]
    \centering
    \includegraphics[width= .7\columnwidth]{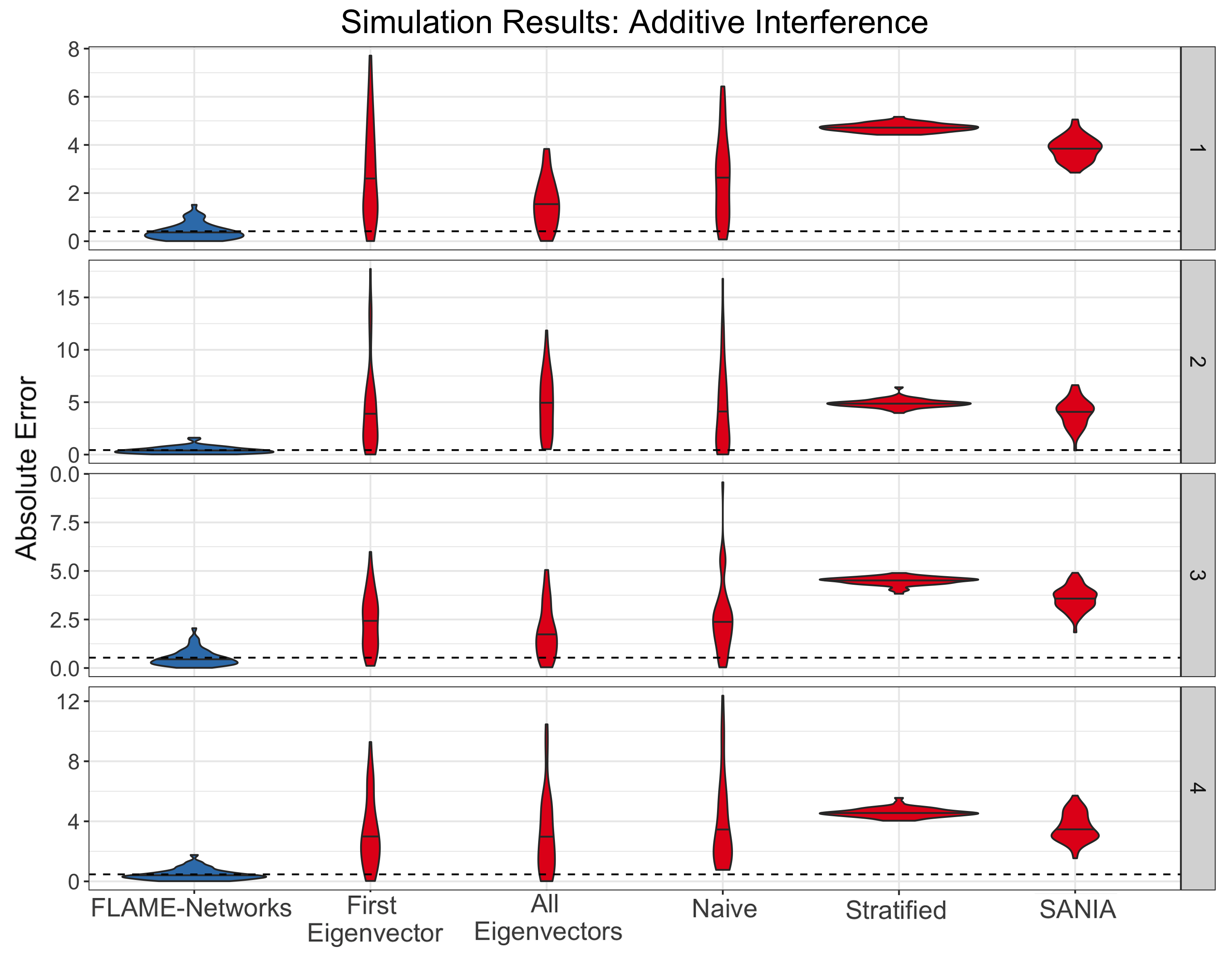}
    \caption{Results from Experiment 1. Each violin plot represents the distribution over simulations of absolute estimation error. The panels are numbered according to the parameter settings of the simulations. Violin plots are blue if the method had mean error lower than or equal to FLAME-Networks' and red otherwise. The black line inside each violin is the median error. The dashed line is FLAME-Networks' mean error.}
    \label{Fig:Additive}
    \vspace{-2mm}
\end{figure}

First we study a setting in which interference is an additive function of the components in Table \ref{Tab:Subgraphs}. Outcomes in this experiment have the form: $Y_i= \gamma_1d_i + \gamma_2\Delta_i + \gamma_3\bigstar^2_i + \gamma_4\bigstar^4_i + \gamma_5\dagger_i^3 + \gamma_6B_i +\gamma_7C_i + \epsilon_i$, with $\epsilon_i \sim N(0, 1)$. We simulate 50 datasets for each setting, in which the units are in an $ER(50, 0.05)$ graph. Table \ref{Tab:AdditiveSettings} shows values for the $\gamma_i$ in each of our experimental settings. 
\begin{table}[]
    \centering
    \begin{tabular}{l|rrrrrrr}
    \hline
         Feature  & $d_i$ & $\Delta_i$ & $\bigstar^2_i$ & $\bigstar^4_i$ & $\dagger^3$ &  $B_i$ & $C_i$\\
         Weight & $\gamma_1$ & $\gamma_2$ & $\gamma_3$ & $\gamma_4$ & $\gamma_5$ &
         $\gamma_6$ & $\gamma_7$\\
         \hline
         Setting 1 & 0 & 10 & 0 & 0 & 0 & 0 & 0 \\
         Setting 2 & 10 & 10 & 0 & 0 & 0 & 0 & 0\\
         Setting 3 & 0 & 10 & 1 & 1 & 1 & 1 & -1\\
         Setting 4 & 5 & 1 & 10 & 1 & 1 & 1 & -1\\
         \hline
    \end{tabular}
    \caption{Settings for Experiment 1.}
    \label{Tab:AdditiveSettings}
    \vspace{-3mm}
\end{table}
Results for Experiment 1 are reported in Figure \ref{Fig:Additive}. FLAME-Networks outperforms all other methods both in terms of average error, and standard deviation over the simulations. This is likely because FLAME-Networks learns weights for the subgraphs that are proportionate to those we use at each setting, and matches units on subgraphs with larger weights. When the interference function is multiplicative instead of additive, FLAME-Networks performs similarly; results are in the appendix.  

\subsection{Experiment 2: Covariate Adjustment}
A strength of FLAME-Networks is its ability to natively account for covariate information. We analyze a setting in which baseline effects are dependent on an additional discrete-valued covariate, $x$, that is observed alongside the network. Outcomes take the form $Y_i = t\tau_i + f(\Gnit) + \beta x_i + \epsilon_i$, where $x_i$ is chosen uniformly at random from  $\{1, 2, 3\}$ for each unit, and $\beta$ is fixed at 15. This means that our sample is divided into 3 strata defined by the covariate values. We ran FLAME-Networks with a dataset consisting of subgraph counts, plus observed covariates for each unit. For comparison with the other methods, we first regress $Y$ on $x$, and then use the residuals of that regression as outcomes for the other methods. This way, the initial regression will account for the baseline effects of $x$, and the residuals contain only potential interference. The interference function takes the form $f(\Gnit) = d_i + \Delta_i +  B_i$, which is what we are trying to learn with the other methods. We simulate the sample network from $ER(50, 0.05)$. 

Results are displayed in Table \ref{Tab:Exp3}. FLAME-Networks performs, on average, better than all the other methods. Results in Section \ref{sec:cov_weight} of the supplement show that when $\beta$ is increased, none of the methods suffer in performance. While regression adjustment prior to estimation seems to have a positive impact on the performance of other methods in the presence of additional covariates, FLAME-Networks performs best. This is because FLAME-Networks is built to easily handle the inclusion of covariates in its estimation procedure. 

\begin{table}[ht]
\centering

\begin{tabular}{lrrr}
  \hline
 Method & Median & 25th q & 75th q \\ 
  \hline
\textbf{FLAME-Networks} & 0.39 & 0.21 & 0.59 \\ 
\hline
  First Eigenvector & 0.47 & 0.40 & 0.83 \\ 
  All Eigenvectors  & 0.55 & 0.29 & 0.79 \\ 
  Naive             & 0.53 & 0.36 & 0.92 \\ 
  SANIA             & 1.93 & 1.75 & 2.25 \\ 
  Stratified        & 4.49 & 4.45 & 4.53 \\ 
  \hline
\end{tabular}
\caption{Results from Experiment 2 with $\beta = 5$. Median and 25th and 75th percentile of absolute error over 40 simulated datasets.}
\label{Tab:Exp3}
\vspace{-2mm}
\end{table}



\subsection{Experiment 3: Misspecified Interference}
We now study the robustness of our estimator in a setting in which one of our key assumptions -- A3 -- is violated. Specifically, we now allow for treated and control units to receive different amounts of interference, \textit{even if they have the same labelled neighborhood graphs}. We do this by temporarily eliminating all control-control edges in the network and then counting the features in Table \ref{Tab:Subgraphs} used to assign interference. That is, consider a unit $i$ with a single, untreated neighbor $j$. In our new setting, if degree is a feature of the interference function, then $i$ being treated implies $i$ receives interference from $j$. But if $i$ is untreated, then $i$ would receive no interference from $j$, because its neighbor is also untreated. This crucially implies that FLAME-Networks will be matching--and estimating the ADE from--individuals that do not necessarily receive similar amounts of interference. 

\begin{figure}[htbp]
    \centering
    \includegraphics[width=.7\columnwidth]{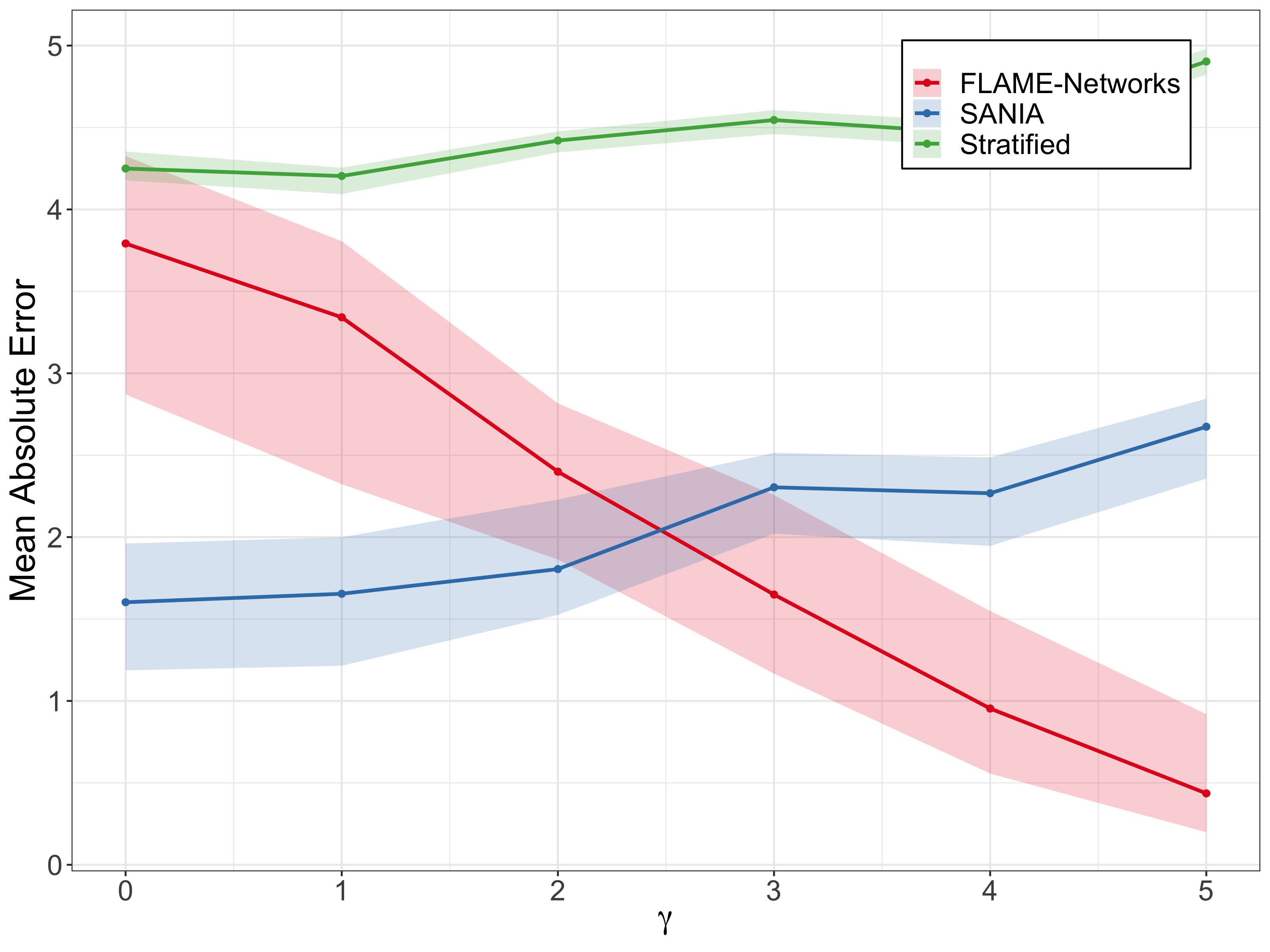}
    \caption{Results from Experiment 3. These simulations were run on an ER(75, 0.07) graph. The bands around the lines represent 25th and 75th quantiles of the 50 simulations, for each value of $\gamma$.}
    \label{Fig:DegVsTri}
    \vspace{-2mm}
\end{figure}

In this setting, we generate interference according to: $f_i = (5 -\gamma) d_i + \gamma\Delta_i$ for $\gamma \in [0, 5]$ and assess the performance of FLAME-Networks against that of the SANIA and stratified estimators. Results are shown in Figure \ref{Fig:DegVsTri}. We see that, when degree is the only component with weight in the interference function, FLAME-Networks performs better than the stratified estimator, but worse than the SANIA estimator, which leverages aspects of the graph related to degree. However, our performance improves as $\gamma$ increases and the true interference depends more on triangle counts since the triangle counts available to FLAME-Networks represent the actual interference pattern more frequently than the degree counts did. Thus, we see that although violation of our method's assumptions harms its performance, it still manages at times to outperform estimators that rely too heavily on degree.

\vspace{-2mm}
\section{APPLICATION}\label{Sec:Application}

In this section, we demonstrate the practical utility of our method.
We use data collected by \cite{banerjee2013diffusion} on social networks for 75 villages in Karnataka, India. They are a median distance of 46 km from one another other, motivating the assumption that network interference is experienced solely between individuals from the same village. 
For each village, we study the effect of 1$.$ lack of education on election participation; 2$.$ lack of education on Self-Help-Group (SHG) participation; and 3$.$ being male on SHG participation. We proxy election participation by ownership of an election card. We compare our estimates -- which account for network interference -- to naive estimates -- which assume no network interference. Data pre-processing is summarized in the appendix. 

\begin{figure}[htbp]
    \centering
    \includegraphics[ width=.5\linewidth]{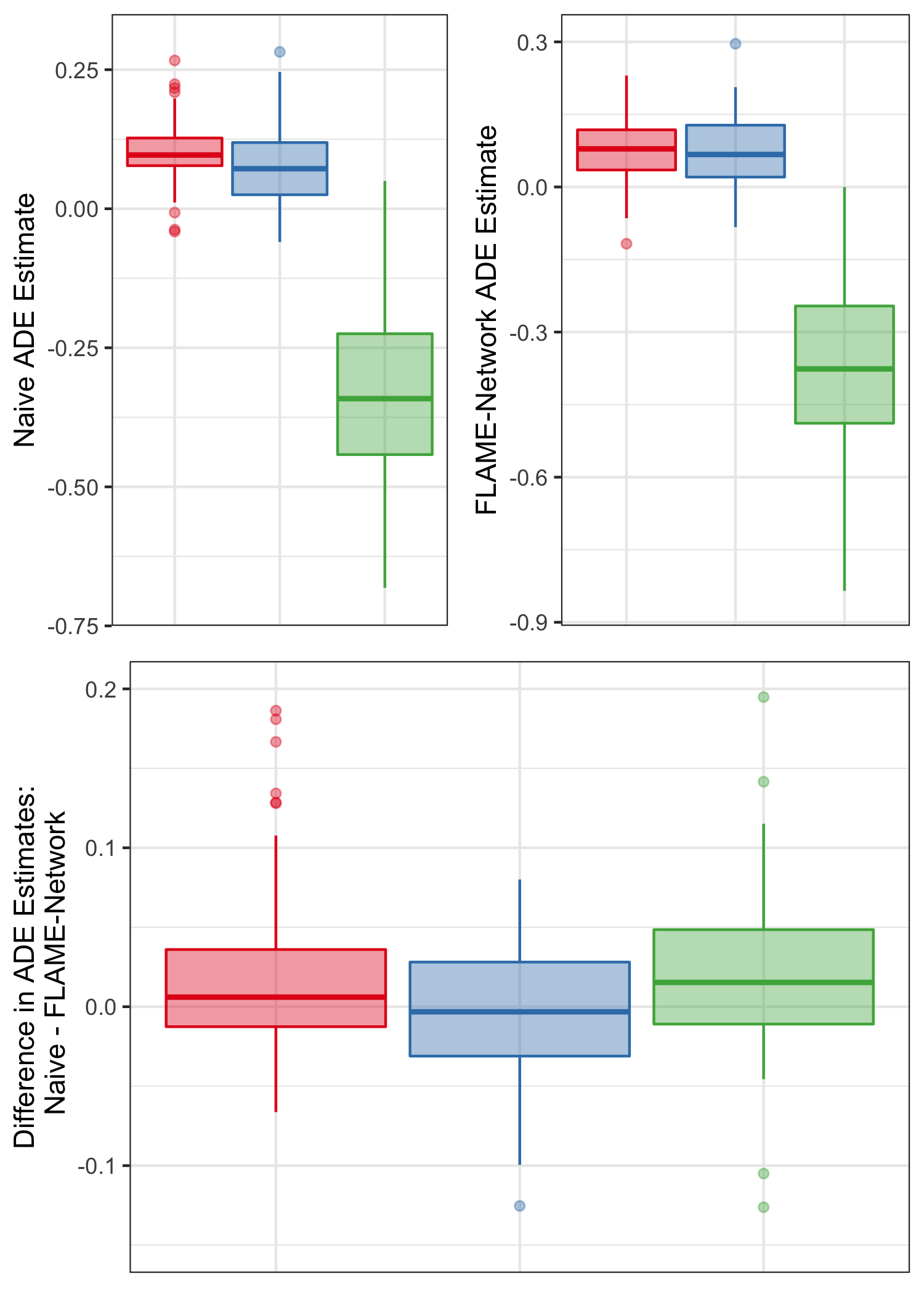}
    \caption{Naive and FLAME-Networks ADE estimates and their difference. Red, blue, and green respectively correspond to (treatment, outcome) pairs: (no education, election participation), (no education, SHG participation), and (gender, SHG participation).}
    \vspace{-2mm}
    \label{Fig:application_diff}
\end{figure}

For ADE estimates, we assume the treatment is randomly assigned. 
We find that lack of education is associated with higher SHG participation, and that males are less likely to participate in SHGs than females (see Figure \ref{Fig:application_diff}). These results make sense in the context of developing countries where SHGs are mainly utilized by females in low-income families, which generally have lower education levels. We observe that education does not impact election participation; in developing countries, an individual's decision to participate in an election may be driven by factors such caste, religion, influence of local leaders and closeness of race \citep{shachar1999follow,gleason2001female}.  FLAME-Networks matches units in each village by subgraph counts and covariate values to estimate the ADE. Looking at the matched groups, we discover that subgraphs such as 2-stars and triangles were important for matching, implying that second-order connections could be affecting interference in this setting. Further details of the matched groups are in Section~\ref{sec:matched_groups}.

Figure \ref{Fig:application_diff} plots naive and FLAME-Networks ADE estimates. We find a significant difference between our estimates and the naive estimates when estimating the effect of being male on participation in SHGs. The naive estimator overestimates the treatment effect, which is possible when ignoring network interference. It is plausible, in this setting, that interference would heighten these effects for all outcomes. This is because individuals from similar social backgrounds or gender tend to interact more together, and, therefore, are more likely to influence each other's participation decision, both in elections and in SHGs. 

%

\vspace{-1mm}
\section{DISCUSSION}
Conventional estimators for treatment effects in randomized experiments will be biased when there is interference between units. We have introduced FLAME-Networks -- a method to recover direct treatment effects in such settings. Our method is based on matching units with similar neighborhood graphs in an almost-exact way, thus producing interpretable, high-quality results. We have shown that FLAME-Networks performs better than existing methods both on simulated data, and we have used real-world data to show how it can be applied in a real setting. Our method extends easily to settings with additional covariate information for units and to taking into account larger neighborhoods for interference.  
In future work, our method can be extended to learning a variety of types of distance metrics between graphs.



\subsubsection*{Acknowledgements}
This work was supported in part by NIH award 1R01EB025021-01, NSF awards IIS-1552538 and IIS1703431, a DARPA award under the L2M program, and a Duke University Energy Initiative ERSF grant.

\newpage
\bibliographystyle{plainnat}
\bibliography{biblio}

\clearpage
\part*{Appendix}
\renewcommand{\thesection}{\Alph{section}}
\newcommand{\pkg}[1]{{\fontseries{b}\selectfont #1}} 
\setcounter{section}{0}
\section{Proofs from the Paper}
\subsection{Proof of Proposition \ref{Thm:Parametric}}
\begin{proof}
We start by writing potential outcomes for an arbitrary unit $i$ from (\ref{eq:additive}) as $Y_i(t, \bt_{-i}) = f_i(t, \btnoti) + \epsilon_i$, where $\epsilon_i$ is some pre-treatment value of the potential outcome, and $f_i(t, \bt_{-i})$ is the treatment response function, dependent both on unit $i$'s treatment and on everyone else's. Using A0-A3 we can write (\ref{eq:additive}) as:
\begin{align}
    Y_i(t, \bt_{-i}) &= t\tau_i + f_i(\bt_{-i}) + \epsilon_i \com{By A1}\nonumber\\
    &= t\tau_i + f_i(\btni) + \epsilon_i  \com{By A2}\nonumber\\
    &= t\tau_i + f(\Gnit) + \epsilon_i \com{By A3}.
\end{align}
This proves Equation \eqref{Eq:OutcomeModel}. The first line agrees with the additivity assumption, which permits a contribution $f_i(\btnoti)$ to $Y_i(t, \btnoti)$ arising from anywhere within the rest of the graph. The second line states that this contribution is limited to $i$'s neighborhood. The third line states that the contribution depends only the neighborhood structure and not anything else about the neighbors.


To prove identification for the ADE, we must show that the individual treatment effect is identified for a treated unit $i$. We need to show that: $\E[Y_i(1, \bzero) - Y_i(0, \bzero)] = \E[Y_i|T_i = 1, \GniT = g_i^\bt] - \E[Y_i|T_i = 0, \GniT = g_i^\bt]$. We have first:
\begin{align}\nonumber
    &\E[Y_i(1, \bzero) - Y_i(0, \bzero)] \\\nonumber
    &= \E[t\tau_i + f(\Gni^{\mathbf{0}}) + \epsilon_i - f(\Gni^{\mathbf{0}}) - \epsilon_i]\\
    &= \E[\epsilon_i + \tau_i - \epsilon_i]
    = \tau_i\label{Eq:ADEPf1}.
\end{align}
Second, we have: 
\begin{align}
    &\E[Y_i|\GniT \simeq g_i^\bt, T_i = 1] - \E[Y_i|\GniT \simeq g_i^\bt, T_i = 0]\nonumber\\
    =& \E[Y_i(1, \bT_{-i})T_i +\nonumber\\
    & \quad + Y_i(0, \bT_{-i})(1-T_i)|\GniT \simeq g_i^\bt, T_i = 1] \nonumber\\
    - &  \E[Y_i(1, \bT_{-i})T_i\nonumber\\
    &\quad + Y_i(0, \bT_{-i})(1-T_i)|\GniT \simeq g_i^\bt, T_i = 0] \nonumber\\
    =& \E[\tau_i + f(g_i^\bt) + \epsilon_i| \GniT \simeq g_i^\bt, T_i = 1] \nonumber\\
    & - \E[f(g_i^\bt) + \epsilon_i|\GnjT = g_i^\bt, T_i = 0]\nonumber\\
    =& \alpha + f(g_i^\bt) + \tau_i - \alpha - f(g_i^\bt)\nonumber\\
    =& \tau_i. \label{Eq:ADEPf2}
\end{align}
The first equality follows from the definition of $Y_i$, the second equality from the result in \eqref{Eq:OutcomeModel} and A3, and the third equality follows from independence of $T$ and $Y$ given in A0: $\E[\epsilon_i|T_i] = \E[\epsilon_i] = \alpha$ for all $i$.
Finally, we can use both of the results above to obtain the ADE:
\begin{align*}
    &\frac{1}{n^{(1)}}\sum_{i=1}^n\E[T_i \times (\E[Y_i|\GniT \simeq g_i^\bt, T_i = 1] \\
    &\qquad\qquad\quad- \E[Y_i|\GniT \simeq g_i^\bt, T_i = 0])]\\
    &= \frac{1}{n^{(1)}}\sum_{i=1}^n\E[T_i\tau_i]\com{By \eqref{Eq:ADEPf2}}\\
    &= \frac{1}{n^{(1)}}\sum_{i=1}^n\E[T_i \times (\E[Y_i(1, \bzero) - Y_i(0, \bzero)])]\com{By \eqref{Eq:ADEPf1}}\\
    &= \frac{1}{n^{(1)}}\sum_{i=1}^n\Pr(T_i = 1)(\E[Y_i(1, \bzero) - Y_i(0, \bzero)])\\
    &= \frac{1}{n^{(1)}}\sum_{i=1}^n\frac{n^{(1)}}{n}(\E[Y_i(1, \bzero) - Y_i(0, \bzero)])\\
    &= \frac{1}{n}\sum_{i=1}^n\E[Y_i(1, \bzero) - Y_i(0, \bzero)],
\end{align*}
where $\Pr(T_i = 1) = \frac{n^{(1)}}{n}$ by assumption of complete randomization.
\end{proof}

\section{Additional Theoretical Results}\label{Sec:Theory}
We study the expected error for one AME match on subgraphs under two assumptions: that the true weights for the AME objective (the weighted Hamming distance) are known, and that the candidate units for matching all have independently generated neighborhoods, and none of the units in these neighborhoods are being matched. Additional information on this setting is available in the proof.  
\begin{proposition}{(Oracle AME Error With Independent Neighborhoods)}\label{Thm:NN}
Suppose that there are $N$  independently generated graphs, each with $n$ vertices, and all i.i.d. from the same distribution over graphs: $\Pr(G_{1} = g_1, \dots, G_{N} = g_N) = \prod_{i=1}^Np(g_i)$. Assume matches are only allowed between units in different graphs. Suppose additionally that $n^{(1)}$ randomly chosen units within each graph are assigned treatment, so that $\Pr(\bT_i = \bt_i) = \binom{n}{n^{(1)}}^{-1}$. Assume further that interference functions obey the following: $|f(g) - f(h)| \leq K \wdist{g}{h}$, where $\bw$ is a vector of positive, real-valued, importance weights for the subgraphs counts, such that $\|\bw\|_1 = M_n$ for some constant $0 < M_n < \infty$, and such that the condition above is satisfied for $\bw$, and $\dist{g}{h}$ is the Hamming distance: a vector of the same length as $\bw$ that is 0 at position $k$ if graphs $g$ and $h$ have the same count of subgraph $k$, and 1 otherwise. Assume that baseline responses have variance $Var(\epsilon_i) = \sigma^2\; \forall i$.  Then, for a treatment unit $i$, if $j$ solves the AME problem, i.e., $j \in \argmin\limits_{\substack{{k =1, \dots, n}\\{T_k = 0}}}\wdist{\hnit}{\GnkT}$, under A0-A3:
\begin{align*}
    &\E\left[|Y_i - Y_j - \tau_i|\big|T_i = 1, \Gnit = \hnit\right] \leq \sqrt{2}\sigma  \\
    &+ K\binom{n-1}{n^{(1)}}^{-1}\times\sum_{g\in \G_n}\sum_{\substack{\bt \in \mathcal{T}\\t_j = 0}}\wdist{h_{\N_i}^{\bt}}{\gnjt}p(g)\\
    &\times \left(\frac{n^{(1)}}{n} + \frac{n - n^{(1)}}{n}C(g_{\N_j}^{\bt})\right)^{N - 2}, 
\end{align*}
where $\G_n$ is the set of all graphs with $n$ units, and $C(h_{\N_j}^{\bt}) \leq 1$ for all $g$ and $\bt$.  
\end{proposition}
A proof is available in the following section. The first element in the right hand side of the inequality is the standard deviation of the baseline responses. One summation is over all possible graphs with $n$ units, and the other summation is over possible treatment assignments. The expression inside the summation is the product of three terms. First, the weighted Hamming distance between a graph and the target graph we are trying to match. Second, the probability of observing that graph. Third, an upper bound on the probability that unit $j$ is among the minimizers of the weighted Hamming distance. 
Note that $\wdist{g_{\N_j}^{\bt}}{\gnjt}p(g)$ is bounded for fixed $n$ for all $g$ and $\bt$. This implies that the bound converges to $2\sqrt{\sigma}$ as $ N \rightarrow \infty$, as long as the size of neighborhood graphs is held fixed, because perfect matching is possible with large amounts of data in this regime. 

\subsection{Proof of Proposition \ref{Thm:NN}}
We briefly review some notation and assumptions to be used in this proof. For the purposes of theory, we study a simplified setting, in which we have to AME-match a unit $i$ to one unit in a set of candidate units of size $N$ such that:
a) all the candidate units belong to disconnected graphs, which we refer to as candidate graphs. b) within each candidate graph there is only one pre-determined candidate unit c) candidate units have neighborhood graphs denoted by $\Gnj$. d) all the candidate graphs are drawn  independently from the same distribution over graphs: $\Pr(G_{\N_1} = g_1, \dots, G_{\N_N} = g_N) = \prod_{i = 1}^Np(g_i)$. The support of $p$ will be $\G_n$: the set of all graphs with exactly $n$ units. We use $g_{\N_i}$ to denote the subgraph induced over $g$ by the units in the set of neighbors of unit $i$, $\N_i \subseteq V(g)$, i.e., $\gni$ is the graph consisting only of the vertices that share an edge with $i$, in $g$, and of the edges in $g$ that are between these vertices. The ego $i$ is not included in $\gni$. 

Assigned treatments are denoted by $\bT$, where $\bT \in \{0, 1\}^n$, but in this setting treatment assignment is assumed to be independent within the $N$ candidate graphs. Formally, the assumption we make is that $\Pr(\bT_1 = \bt_1, \dots, \bT_N = \bt_N) = \prod_{i=1}^N\binom{n}{n^{(1)}}^{-1}$, i.e., $n^{(1)}$ units are always treated uniformly at random within each of the $N$ candidate graphs. 

The direct treatment effect for any unit $i$ is given by $\tau_i$.  We use $\dist{g}{h}$ to indicate the Hamming distance between subgraph counts of graphs $g$ and $h$. This means that $\dist{g}{h}$ is a vector of size $|\G_n|$ that will be 1 in the $\ell^{th}$ entry if $g$ and $h$ have the same amount of occurrences of graph $g_\ell$ among their subgraphs. Note that this distance is coloring sensitive: two subgraphs that are isomorphic in shape but not labels will belong to different entries in this distance. The matched group of a \textit{treated} unit $i$, denoted $\MG_i$ is the set of all \textit{control} units that match with $i$. In our setting $j \in \MG_i$ if it solves the AME problem, that is $j \in \argmin\limits_{\substack{k=1\dots, n\\T_k = 0}}\wdist{\GnkT}{\gnit}$. Finally, we assume that both the graph for the unit we want to match and the treatment assignment for that unit's graph are fixed: $\bt_i$ is the treatment assignment in the graph of $i$, and $\hniti$ is the neighborhood graph of $i$, where $h$ denotes unit $i$'s graph. All other notation is as in the main paper.

\begin{proof}
We start by upper-bounding our quantity of interest as follows:
\begin{align}
    &\E[|Y_i - Y_j - \tau_i|\big|j \in \MG_i]\nonumber \\
    &= \E[|Y_i(1, \bt_{i-i}) - Y_j(0, \bT_{j-j}) - \tau_i|\big|j \in \MG_i]\nonumber\\
    &= \E[|\tau_i + f(\hniti) + \epsilon_i - f(\GnjTj) - \epsilon_j - \tau_i|\big|j \in \MG_i]\nonumber\\
    &\leq \E[|f(\hniti) - f(\GnjTj)|\big|j \in \MG_i] \nonumber\\
    & \quad +  \E[| \epsilon_i - \epsilon_j|\big|j \in \MG_i]\nonumber\\
    &\leq K\E[\wdist{\hniti}{\GnjTj}\big|j \in \MG_i]\nonumber \\
    &\quad + \E[| \epsilon_i - \epsilon_j|\big|j \in \MG_i],\label{Eq:BiasDecomp1}
\end{align}
where the notation $\bT_{j-j}$ denotes the treatment indicator for candidate graph $j$ for all units except $j$. The first equality follows from A0 since the event $j \in \MG_i$ implies that $T_j = 0$, as only control units are allowed in the matched groups. The second equality follows from Proposition \ref{Thm:Parametric}. The first inequality is an application of the triangle inequality. The last line follows from the condition on the interference functions.  
Consider the second term. We can use the Cauchy-Schwarz inequality to construct a simple upper bound on it:
\begin{align*}
    &\E[|\epsilon_i - \epsilon_j|\big|j \in \MG_i] = \E[|\epsilon_i - \epsilon_j|]\\
    &\leq \sqrt{\E[(\epsilon_i - \epsilon_j)^2]} \\
    &= \sqrt{\E[\epsilon_i^2] + \E[\epsilon_j^2] - 2\E[\epsilon_i]\E[\epsilon_j]} = \sqrt{2}\sigma
\end{align*}
where the last equality follows for the fact that the $\epsilon_i$ have mean 0 and are independent, with $Var(\epsilon_i) = \sigma^2$ for all $i$.

Consider now the term $\E[\wdist{\hniti}{\GnjTj}|j \in \MG_i]$. To upper-bound this, we write it out as follows using the definition of expectation:
\begin{align*}
    &\E[\wdist{\hniti}{\GnjTj}|j \in \MG_i] \\
    &= \sum_{g \in \G_n}\sum_{\bt \in \cT, t_j = 0}\wdist{\hniti}{\gnjtj}\\
    &\qquad\qquad\times \Pr(\GnjTj = \gnjt, \bT_j=\bt|j \in \MG_i). 
\end{align*}
We want to find an upper bound on $\Pr(\GnjTj = \gnjt, \bT_j=\bt|j \in \MG_i)$. We start by writing this quantity out as a product of two probabilities:
\begin{align*}
    &\Pr(\GnjTj = \gnjt, \bT_j=\bt|j \in \MG_i) \\
    &= \Pr(\GnjTj = \gnjt|j \in \MG_i, \bT_j = \bt)\\
    &\times\Pr(\bT_j=\bt|j \in \MG_i)\\
    &= \Pr(\GnjTj = \gnjt|j \in \MG_i, \bT_j = \bt)\binom{n-1}{n^{(1)}}^{-1}.
\end{align*}
Note that $\Pr(\bT_j=\bt|j \in \MG_i) = \binom{n-1}{n^{(1)}}^{-1}$ because treatment is uniformly randomized with $n^{(1)}$ units always treated in each candidate graph, but $T_j = 0$ conditionally on $j \in \MG_i$. 

We use Bayes' rule to write out the first term in the final product as $\Pr(\GnjT = \gnjt|j \in \MG_i, \bT_j = \bt) = \frac{\Pr(j \in \MG_i|\bT_j = \bt, \GnjTj = \gnjt)\Pr(\GnjTj = \gnjt|\bT_j = \bt)}{\Pr(j \in \MG_i|\bT_j = \bt)}$. 

By assumption, if all neighborhood graphs are empty, all units are used for all matched groups, and we are restricting ourselves to assignments in which $T_j=0$, therefore, $\Pr(j \in \MG_i|\bT_j = \bt) = 1$. Second, by assumption $\Pr(\GnjTj = \gnjt|\bT_j = \bt) = p(g)$. This is because treatment assignment is independent of the graph. We are left with having to find an expression for the likelihood, this can be written as:
\begin{align*}
    &\Pr(j \in \MG_i|\bT_j = \bt, \GnjTj = \gnjt)\\
    &= \Pr(j \in \argmin_{\substack{k = 1, \dots, N, \\k\neq i}} \wdist{\hniti}{\GnkTk}\\
    &\qquad\qquad|\bT_j = \bt, \GnjT = \gnjt)\\
    &= \prod_{\substack{k = 1\\k\neq i, j}}^N \bigg[\Pr(\wdist{\hniti}{\GnkTk} \geq \\ 
    &\qquad \qquad \: \quad \wdist{\hniti}{\gnjt}| T_k = 0)\Pr(T_k = 0) \\
    &\qquad\quad + \Pr(T_k = 1)\bigg]\\
    &= \prod_{\substack{k = 1\\k\neq i, j}}^N \bigg[\Pr\big(\wdist{\hniti}{\GnkTk} \\
    &\qquad\quad\ \ \,\geq \wdist{\hniti}{\gnjt}| T_k = 0\big)\frac{n - n^{(1)}}{n}\\
    &\qquad\quad+\frac{n^{(1)}}{n}\biggl].
\end{align*}
The second equality follows because $k$ can never be in the matched group of unit $i$ if $T_k=1$, and, if $T_k=0$, then $k$ must be one of the minimizers of the weighted Hamming distance between neighborhood subgraph counts. The probability is a product of densities because of independence of candidate subgraphs. For an arbitrary unit, $k$, we define the following compact notation for the probability that $k$'s weighted Hamming distance from $i$ is larger than the weighted Hamming distance from $j$ to $i$:
\begin{align*}
    &\Pr(\wdist{\hniti}{\GnkTk}\\
    &\qquad\geq \wdist{\hniti}{\gnjt}| T_k = 0)\\
    &=: C_k(\gnjt) \leq 1. 
\end{align*}
Note that the last inequality follows from the fact that the expression above is a probability. Since graphs and treatment assignments, $G_k$ and $\bT_k$ are the only random variables in the probability denoted by $C_k(\gnjt)$, and since they are all independent, and identically distributed, we can say that $C_1(\gnjt) = C_2(\gnjt) = \dots = C_N(\gnjt) = C(\gnjt)$. Because of this we have:
\begin{align*}
    &\Pr(j \in \MG_i|\GnjTj = \gnjt, \bT_j = \bt) \\
    &= \left(\frac{n^{(1)}}{n} + \frac{n - n^{(1)}}{n}C(\gnjt)\right)^{N - 2}.
\end{align*}
Putting all the elements we have together we get the expression for the first term in the bound:
\begin{align*}
    &\E[\wdist{\hniti}{\GnjTj}|j \in \MG_i] \\
    &= \sum_{g \in \G_n}\sum_{\bt \in \cT:
    t_j = 0}\wdist{\gnjt}{\hniti}\\
    &\qquad\qquad\times \Pr(\GnjT = \gnjt, \bTnj=\btnj|j \in \MG_i)\\
    &= \sum_{g \in \G_N}\sum_{\bt \in \cT: t_j = 0}\wdist{\gnjt}{\hniti}\\
    &\times \binom{n-1}{n^{(1)}}^{-1}p(g)\left(\frac{n^{(1)}}{n} + \frac{n - n^{(1)}}{n}C(\gnjt)\right)^{N - 2}.
\end{align*}
\end{proof}

\subsection{Asymptotic Behavior}
Here we expand on the asymptotic consequences of Proposition \ref{Thm:NN}: note first, that, by assumption $\|\bw\|_1 = M_n$, and that, therefore $\wdist{g}{h}\leq M_n$ for any graphs $g, h \in \G_n$. That is to say, the weighted Hamming distance between any two graphs with $n$ units will be upper-bounded by the sum of the weights. Recall also that $C(\gnjt) \leq 1$ for all $g$ and $\bt$ as this quantity is a probability, and let $\Cmax = \max\limits_{\substack{g \in \G_n\\\bt \in \cT}}C(\gnjt)$. We can combine all these bounds with the upper bound in Proposition \ref{Thm:NN} to write:
\begin{align*}
    &\E[\wdist{\hniti}{\GnjTj}|j \in \MG_i] \\
    &= \sum_{g \in \G_\N}\sum_{\bt \in \cT, t_j = 0}\wdist{\gnjt}{\hniti}\\
    &\times \binom{n-1}{n^{(1)}}^{-1}p(g)\left(\frac{n^{(1)}}{n} + \frac{n - n^{(1)}}{n}C(\gnjt)\right)^{N - 2}\\
    &\leq M_n\left(\frac{n^{(1)}}{n} + \frac{n - n^{(1)}}{n}\Cmax\right)^{N - 2}\\
    &\times \sum_{g \in \G_n}\sum_{\bt \in \cT, t_j = 0}\binom{n-1}{n^{(1)}}^{-1}p(g)\\
    &= M_n\left(\frac{n^{(1)}}{n} + \frac{n - n^{(1)}}{n}\Cmax\right)^{N - 2}. 
\end{align*}
The first equality follows from Proposition \ref{Thm:NN}, the first inequality from the bounds previously discussed, and the second equality follows from the fact that the sum in the second to last line is a sum of probability distributions over their entire domain, and therefore is equal to 1. Under the condition that $n$, the number of units in each unit's candidate graph, stays fixed, and that $\Cmax < 1$, then, as $N\rightarrow \infty$, we have $M_n\left(\frac{n^{(1)}}{n} + \frac{n - n^{(1)}}{n}\Cmax\right)^{N - 2} \rightarrow 0$, because the quantity inside the parentheses is always less than 1. This makes sense, because asymptotically, matches can be made exactly; i.e., units matched in the way described in our theoretical setting have isomorphic neighborhood subgraphs asymptotically. This also has a consequence that the bound in Proposition \ref{Thm:NN} converges to $\sqrt{2}\sigma$ asymptotically in $N$. This is the variance of the baseline errors and can be lowered by matching the same unit with multiple others. As noted before, for this argument to apply, candidate graphs must remain of fixed size $n$ as they grow in number, so that the quantity $M_n$ remains constant: this setting is common in cluster-randomized trials where a growing number of units is clustered into fixed-size clusters of size at most $n$. The asymptotic behavior of our proposed methodology is less clear in settings in which the analyst cannot perform such clustering before randomization and $n$ is allowed to grow with $N$, and is an avenue for potential future research.  

\subsection{Heteroskedasticity in The Baseline Effects}
In a network setting such as the one we study, it is possible that baseline effects of units do not have equal variance. Here we discuss how this setting affects our result in Proposition \ref{Thm:NN}. Here, we maintain that $\E[\epsilon_i]=\alpha$ for all $i$, but we assume that $Var(\epsilon_i) = \sigma_i^2$, and that $Cov(\epsilon_i\epsilon_j)\neq 0$. Starting from the upper bound on the estimation error given in \eqref{Eq:BiasDecomp1}, we can see that the baseline effects only come in in the term: $\E[|\epsilon_i - \epsilon_j||j \in \MG_i]$, we therefore focus our attention on this term, as the rest of this bound does not change when the variance of these terms changes. Note first, that $\E[|\epsilon_i - \epsilon_j||j \in \MG_i] = \E[|\epsilon_i - \epsilon_j|]$ as the event $j \in \MG_i$ is independent of the baseline effects. We can now apply the Cauchy-Schwarz inequality, in the same way as we do in the proof of Proposition \ref{Thm:NN}, to obtain:
\begin{align*}
    \E[|\epsilon_i - \epsilon_j|] &\leq \sqrt{\E[(\epsilon_i - \epsilon_j)^2]} \\
    &= \sqrt{\E[\epsilon_i^2] + \E[\epsilon_j^2] - 2\E[\epsilon_i]\E[\epsilon_j]}\\
    &= \sqrt{\sigma^2_i + \alpha^2 + \sigma_j^2 + \alpha^2 - 2\alpha^2}\\
    &= \sqrt{\sigma_i^2 + \sigma_j^2}. 
\end{align*}
Clearly, this is not too different from the homoskedastic setting we study in the proposition: as long as neither of the unit variances is too large for inference, results in the heteroskedastic setting will suffer from similar bias as they would under\textbf{} independent baseline effects with equal variance.

Simulations, shown in Section \ref{sec:heteronetwork}, also support the above rationale of comparable performance in the heteroskedastic case and demonstrate that FLAME-Networks still outperforms competing methods.

\clearpage

\section{Derivation of SANIA MIVLUE Used in Simulations}

\textbf{Theorem 6.2 of Sussman and Airoldi, 2017}\\
\textit{Suppose potential outcomes satisfy SANIA and that the prior on the parameters (baseline outcome and direct treatment effect) has no correlation between units. If unbiased estimators exist, the MIVLUE weights are:}
\[w_i(\mathbf{z}) = \frac{C_{i, d_i^{\mathbf{z}}}}{\sum_{d=0}^{n-1}C_{i, d}} \cdot \frac{2z_i - 1}{n\mathrm{P}(\mathbf{z}_i^{\text{obs}} = z_i, d_i^{\mathbf{z}^{\text{obs}}}=d_i^{\mathbf{z}})}\]
where 
\begin{align*}C_{i, d} &= \\ &\left(\sum_{\mathbf{z}} \mathrm{P}(\mathbf{z})\mathbf{1}\{d_i^{\mathbf{z}} = d\} \cdot \frac{\Sigma(\mathbf{z})_{ii}}{\mathrm{P}(z_i^{\text{obs}} = z_i, d_i^{\mathbf{z}^{\text{obs}}} = d)^2}\right)^{-1}\end{align*}
and $C_{i, d}$ is defined to be 0 if the probability in its denominator is 0.

\subsection{Setup and Notation}

We assume that there is a constant probability $p$ of each unit being treated and that units are independently treated. Let unit $i$ have $d_i$ neighbors and a treated degree of $d_i^{\mathbf{z}}$. In our setting, $\Sigma(\mathbf{z})_{ii} = \Sigma_{\alpha, ii} + z_i\Sigma_{\beta, ii}$ where $\Sigma_{\alpha}$ and $\Sigma_{\beta}$ are the covariance matrices on priors placed on the baseline outcome and the direct treatment effect, respectively. Additionally in our setting, their diagonals are constant and so we let $\sigma^2_{\alpha} := \Sigma_{\alpha, ii}$ and $\sigma^2_{\beta} := \Sigma_{\beta, ii}$.

\subsection{Find $\mathrm{P}(z_i^{\text{obs}} = z_i, d_i^{\mathbf{z}^{\text{obs}}} = d_i^{\mathbf{z}})$}
By the setup, the constituent probabilities are independent and the probability of treatment is constant across units and so: $\mathrm{P}(z_i^{\text{obs}} = z_i, d_i^{\mathbf{z}^{\text{obs}}} = d_i^{\mathbf{z}}) = [z_i p + (1 - z_i) (1 - p)]{d_i \choose d_i^{\mathbf{z}}}p^{d_i^{\mathbf{z}}}(1 - p)^{d_i - d_i^{\mathbf{z}}}$

\subsection{Find $C_{i, d}$}
Below, we will consider $\mathbf{z}_{\text{neighbor}(i)}$, the treatment assignment of the neighbors of $i$ (excluding $i$), $z_i$, the treatment assignment of unit $i$, and $\mathbf{z}_{\text{rest}(i)}$, the treatment assignment of the remaining units. 
\begin{align*}
	C_{i, d} &= \left(\sum_{\mathbf{z}} \frac{\mathrm{P}(\mathbf{z})\mathbf{1}\{d_i^{\mathbf{z}} = d\}\Sigma(\mathbf{z})_{ii}}{\mathrm{P}(z_i^{\text{obs}} = z_i, d_i^{\mathbf{z}^{\text{obs}}} = d)^2}\right)^{-1} \\
	&= \left(\sum_{\mathbf{z}: d_i^{\mathbf{z}}=d} \frac{\mathrm{P}(\mathbf{z}_{\text{neighbor}(i)})\mathrm{P}(z_i)\mathrm{P}(\mathbf{z}_{\text{rest}(i)})\Sigma(\mathbf{z})_{ii}}{\mathrm{P}(z_i^{\text{obs}} = z_i, d_i^{\mathbf{z}^{\text{obs}}} = d)^2}\right)^{-1} \\
	&= \left(\frac{\mathrm{P}(\mathbf{z}_{\text{neighbor}(i)})}{\mathrm{P}(z_i^{\text{obs}} = z_i, d_i^{\mathbf{z}^{\text{obs}}} = d)^2}\right)^{-1} \\
	&\quad\times \left(\sum_{\mathbf{z}: d_i^{\mathbf{z}}=d} \Sigma(\mathbf{z})_{ii}\mathrm{P}(z_i)\mathrm{P}(\mathbf{z}_{\text{rest}(i)})\right)^{-1} \\
	&= \left(\frac{\mathrm{P}(\mathbf{z}_{\text{neighbor}(i)})}{\mathrm{P}(z_i^{\text{obs}} = z_i, d_i^{\mathbf{z}^{\text{obs}}} = d)^2}\right)^{-1} \\
	&\quad\times \left(\sum_{\mathbf{z}: d_i^{\mathbf{z}}=d} (\sigma^2_{\alpha} + z_i\sigma^2_{\beta})\mathrm{P}(z_i)\mathrm{P}(\mathbf{z}_{\text{rest}(i)})\right)^{-1}\\
	&= \left(\frac{\mathrm{P}(\mathbf{z}_{\text{neighbor}(i)})}{\mathrm{P}(z_i^{\text{obs}} = z_i, d_i^{\mathbf{z}^{\text{obs}}} = d)^2}\right)^{-1} \\
	&\quad\times\left(\sigma^2_{\alpha} + \sum_{\mathbf{z}: d_i^{\mathbf{z}}=d} (z_i\sigma^2_{\beta})\mathrm{P}(z_i)\mathrm{P}(\mathbf{z}_{\text{rest}(i)})\right)^{-1}\\
	&= \left(\frac{\mathrm{P}(\mathbf{z}_{\text{neighbor}(i)})}{\mathrm{P}(z_i^{\text{obs}} = z_i, d_i^{\mathbf{z}^{\text{obs}}} = d)^2}\right)\\
	&\quad\times\left(\sigma^2{\alpha} + \sum_{\substack{\mathbf{z}: d_i^{\mathbf{z}}=d\\z_i = 1}}(\sigma^2_{\beta,})p\mathrm{P}(\mathbf{z}_{\text{rest}(i)})\right)^{-1} \\
	&= \left(\frac{\mathrm{P}(\mathbf{z}_{\text{neighbor}(i)})\left(\sigma^2_{\alpha} + \sigma^2_{\beta}\right)}{\mathrm{P}(z_i^{\text{obs}} = z_i, d_i^{\mathbf{z}^{\text{obs}}} = d)^2}\right)^{-1}\\
	&= \frac{\left([z_i p + (1 - z_i) (1 - p)]{d_i \choose d}p^{d}(1 - p)^{d_i - d}\right)^2}{(\sigma^2_{\alpha} + \sigma^2_{\beta})p^d(1-p)^{d_i - d}} \\
	&= \frac{[z_i p + (1 - z_i) (1 - p)]^2{d_i \choose d}^2p^{d}(1 - p)^{d_i - d}}{\sigma^2_{\alpha} + \sigma^2_{\beta}} \\
\end{align*}

\subsection{Find $w_i$}
Plugging in the expressions we've found:
\begin{align*}
	&w_i(\mathbf{z}) = \\
	&\frac{C_{i, d_i^{\mathbf{z}}}}{\sum_{d=0}^{n-1}C_{i, d}} \cdot \frac{2z_i - 1}{n\mathrm{P}(\mathbf{z}_i^{\text{obs}} = z_i, d_i^{\mathbf{z}^{\text{obs}}}=d_i^{\mathbf{z}})} \\
	&= \frac{\frac{[z_i p + (1 - z_i) (1 - p)]^2{d_i \choose d_i^{\mathbf{z}}}^2p^{d_i^{\mathbf{z}}}(1 - p)^{d_i - d_i^{\mathbf{z}}}}{(\sigma^2_{\alpha} + \sigma^2_{\beta})}}{\frac{\sum_{d=0}^{n-1}{d_i \choose d}^2p^d(1-p)^{d_i - d}(z_i p + (1 - z_i)(1 - p))^2}{(\sigma^2_{\alpha} + \sigma^2_{\beta})}}\\
	&\quad\times \frac{2z_i - 1}{n[z_i p + (1 - z_i) (1 - p)]{d_i \choose d_i^{\mathbf{z}}}p^{d_i^{\mathbf{z}}}(1 - p)^{d_i - d_i^{\mathbf{z}}}} \\
	&= \frac{(z_i p + (1 - z_i) (1 - p))(2z_i - 1)}{n\sum_{d=0}^{n-1}{d_i \choose d}^2p^d(1-p)^{d_i - d}(z_i p + (1 - z_i)(1 - p))^2} \\
	&= \frac{(z_i p + (1 - z_i) (1 - p))(2z_i - 1)}{n(z_i p + (1 - z_i)(1 - p))^2}\\
	&\quad\times\frac{1}{\sum_{d=0}^{n-1}{d_i \choose d}^2p^d(1-p)^{d_i - d}}
\end{align*}
Note in the first fraction that $(z_i p + (1 - z_i) (1 - p))(2z_i - 1)$ equals $p$ when $z_i = 1$ and $p - 1$ when $z_i = 0$. Also, $(z_i p + (1 - z_i) ( 1 - p))^2$ equals $p^2$ when $z_i = 1$ and $(1 - p)^2$ when $z_i = 0$. Thus, the first term is $1/np$ when $z_i = 1$ and $-1/n(1 - p)$ when $z_i = 0$ and so the overall expression for the weights is:
\[w_i(\mathbf{z}) = \frac{z_i/np - (1 - z_i)/(n(1-p))}{\sum_{d=0}^{n-1}{d_i \choose d}^2p^d(1-p)^{d_i - d}}
\]
and the MIVLUE is given by $\sum_{i=1}^n w_i Y_i$.

\section{Subgraph Descriptions}
\begin{figure}[]
    \centering
    \includegraphics[ height=1\linewidth]{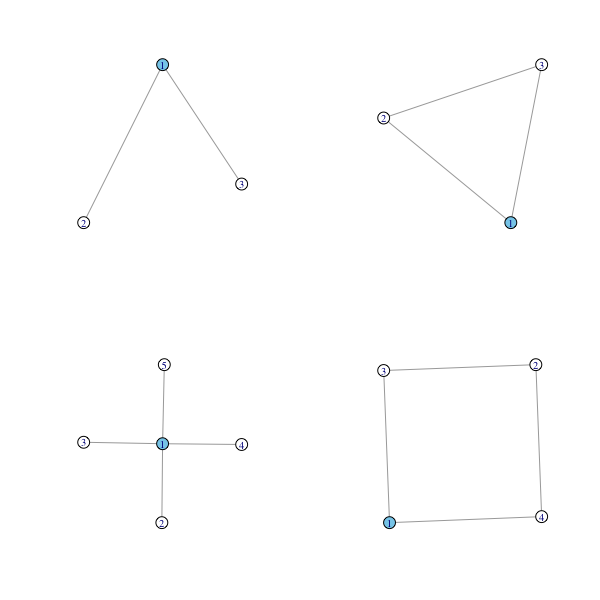}
    \caption{Four types of treated neighborhood subgraphs the colored unit might be a part of: triangle (top right), square (bottom right), 2-star (top left), 4-star (bottom left).}
    \label{Fig:Subgraph_Ex}
\end{figure}

Here, for graphs without self-loops, we define the interference components used in our simulations:
\begin{itemize}
    \item Degree: the degree of a node is the number of edges it is a part of. \\
    \item Triangle: A graph with three mutually connected nodes (see Figure \ref{Fig:Subgraph_Ex}). \\
    \item Square: A graph with 4 nodes and 4 edges, such that each node is a part of exactly two distinct edges (see Figure \ref{Fig:Subgraph_Ex}). \\
    \item $k$-Star: A graph with $k + 1$ nodes, the first $k$ of which are all connected to the $(k + 1)$st node and no others (see Figure \ref{Fig:Subgraph_Ex}).
    \item Vertex Betweenness: The vertex betweenness of a vertex $v$ is defined as:
    \[\sum_{i \neq v \neq j} \frac{\sigma_{ij}(v)}{\sigma_{ij}}\]
    where $\sigma_{ij}$ is the number of shortest paths between vertices $i$ and $j$ and $\sigma_{ij}(v)$ is the number of shortest paths between $i$ and $j$ that go through $v$. We use the normalized vertex betweenness which scales the above expression by $2/(n^2 - 3n + 2)$ where $n$ is the number of nodes in the graph. \\
    \item Closeness Centrality: We use the normalized closeness centrality of a vertex $v$, defined as:
    \[\frac{n-1}{\sum_{i=1}^n d(\sigma_{vi})}\]
    where $d(\sigma_{vi})$ is the length of the shortest path between $v$ and $i$ and $n$ is the number of nodes in the graph. 
\end{itemize}

\section{Data Pre-processing}
We estimate the DTE using data from from \cite{banerjee2013diffusion} on social networks for the 75 villages in India. A unit $i$ is defined to be socially connected with unit $j$ if they are connected across at least 3 of the following four types of social connections: (1) $i$ visits $j$'s house, (2) $j$ visits $i$'s house, (3) $i$ and $j$ socialize and are relatives, (4) $i$ and $j$ socialize and are not related to each other. Along with subgraph counts, we also use age and whether or not individual $i$ speaks Telugu as additional covariates to match on. For computational efficiency, we drop units with maximum degree of connection greater than 15, where the cut-off is selected based on computational efficiency.

\section{Matched Groups}
\label{sec:matched_groups}
We provide sample matched groups in Table~\ref{tab:samplegroups}. These matched groups were produced by applying FLAME-Networks on the data discussed in Section \ref{Sec:Application}. We report all the covariates used for matching. The first group is comprised of 40-year-old units who do not speak Telugu, and have 2 or 3 triangles, 3 2-stars, and 3 edges in their treated neighborhood graph. These units (given the binning) are matched exactly. The second group is comprised of units who speak Telugu with no triangles, 3 2-stars, 3 edges in their treated neighborhood graph. Note that units in this group are matched approximately, since they are not matched exactly on age.

\begin{table}[]
\centering
\begin{minipage}{0.5\textwidth}
\resizebox{\textwidth}{!}{%
\begin{tabular}{llllllrr}
\hline
Units & Triangles  & 2-Stars  &  Edges & Telugu & Age  &  Treated &  Outcome \\
\hline
Matched Group 1\\
\hline
1 &         2 or 3 &  3 &  3 &  0 &       40 &        0 &        1 \\
2 &         2 or 3 &  3 &  3 &  0 &   40 &        1 &        0 \\
3 &         2 or 3 &  3 &  3 &  0 &   40 &        1 &        0 \\
4 &         2 or 3 &  3 &  3 &  0 &   40 &        1 &        0 \\
\hline
Matched Group 2\\
\hline
1 &         0 &  3 &  3 &  1 &       30 &        0 &        0 \\
2 &         0 &  3 &    3 &  1 &   34 &        0 &        0 \\
3 &         0 &  3 &    3 &  1 &   25 &        0 &        1 \\
4 &         0 &  3 &    3 &  1 &   20 &        1 &        0 \\
\hline
\end{tabular}
}
\caption{Sample Match Groups. Two sample matched groups generated by FLAME-Networks using data discussed in Section \ref{Sec:Application}. The columns are the covariates used for matching, along with treatment status and outcome. The counts of subgraphs were coarsened into 10 bins defined by deciles of the empirical distribution of counts. The two groups have relatively good match quality overall. Note that the first group matches units exactly (given the binning). However, Matched Group 2 matches units approximately, with exact matches on subgraph counts and whether or not individuals speak Telugu, but inexact matches on age.}
\label{tab:samplegroups}
\end{minipage}
\end{table}

\section{Additional Experiment: Multiplicative Interference}
We explore settings in which interference is a nonlinear function of the interference components and their weights. Since matching is nonparametric, it is particularly appropriate for handling non-linearities in interference functions. Outcomes in this experiment have the form $Y_i(t, \btnoti) = t\tau_i + \alpha\prod_{p=1}^Pm_{ip}^{\ind[p \text{ is included}]} + \epsilon_i$. Table \ref{Tab:SettingsMultiplicative} shows which components are included in the outcome function for each setting. 
We use a small number of parameters in each setting, as their multiplicative interaction suffices to define a complex interference function. The simulations are run on an $ER(50, 0.05)$ graph.

\begin{table}[htbp]
    \centering
    \resizebox{0.6\columnwidth}{!}{%
    \begin{tabular}{r|cccc}
    \hline
     Setting & $d_i$ & $\Delta_i$ & $\bigstar_i^4$ & $B_i$ \\ 
     \hline
        1    & x & x & &   \\
        2    & x &   & & x \\
        3    &   & x & & x \\
        4    &   & x & x & \\
        \hline
    \end{tabular}}
    \caption{Parameters included in interference function Experiment 1. The marked components for each setting were the only ones included in those experiments.}
    \label{Tab:SettingsMultiplicative}
\end{table}

\begin{figure}[!htbp]
    \centering  
    \includegraphics[width=.99\linewidth]{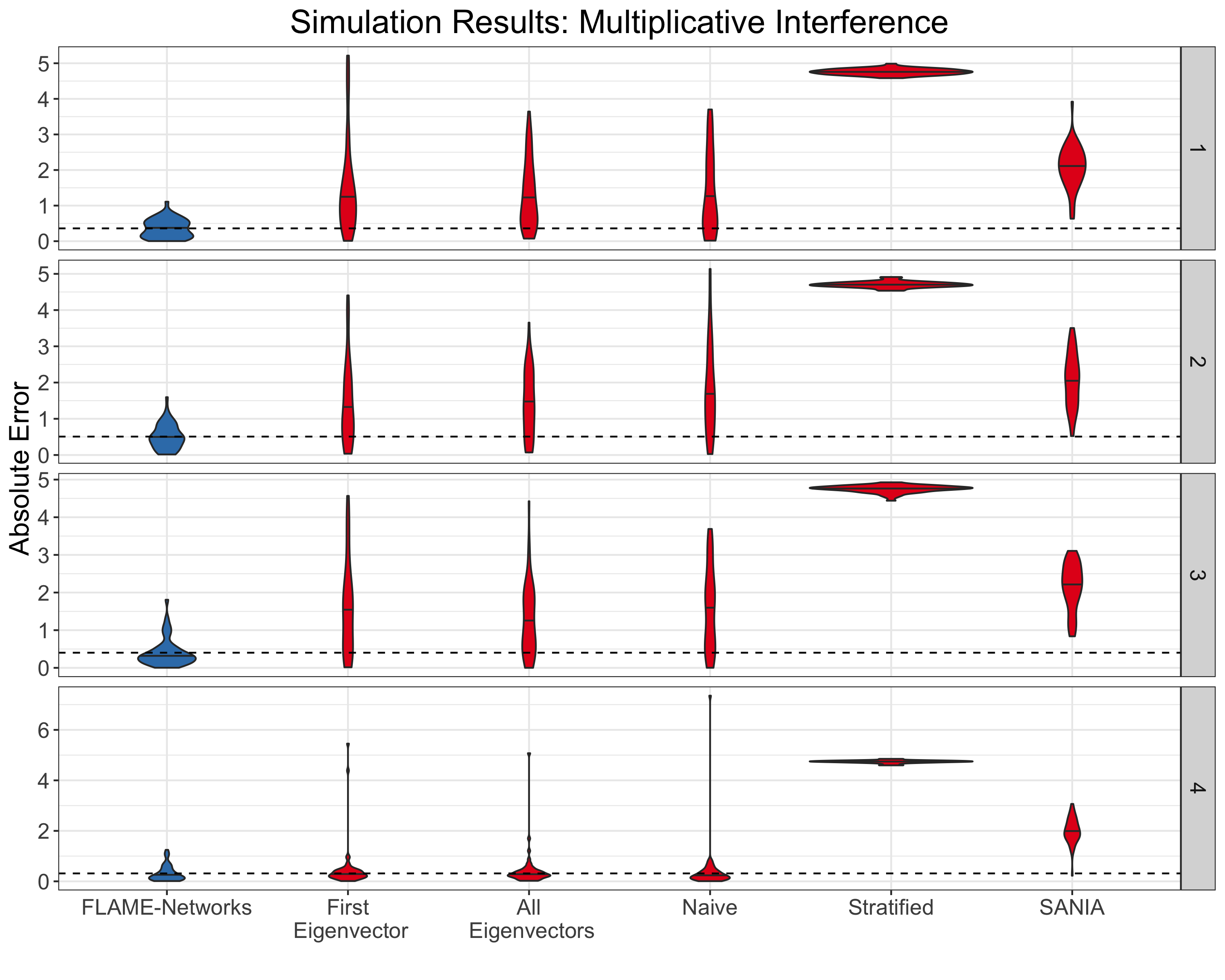}
    \caption{Results from experiments with a multiplicative interference function.  Each violin plot represents the distribution over simulations of absolute estimation error over for each method. The panels are numbered according to the parameter setting the simulations were ran with. Violin plots are color-coded blue if the method had mean error either equal to or better than FLAME-Networks and red otherwise. The black line inside each violin is the median error. The dashed line is FLAME-Networks' mean error.}
    \label{Fig:Multiplicative}
\end{figure}

Results for this experiment are presented in Figure \ref{Fig:Multiplicative}. FLAME-Networks performed better than all baseline methods in this setting, both in terms of mean absolute error and, in most cases, in terms of standard deviation over simulations. The stratified and SANIA estimators perform especially poorly, because they cannot handle nonlinear interference settings, unlike FLAME-Networks. 
\vspace{-1mm}

\section{Additional Experiment: Graph Cluster Randomization}\label{sec:graphcluster}
\begin{figure}[!htbp]
    \centering  
    \includegraphics[trim={0 0 0 2cm}, clip, width=.99\linewidth]{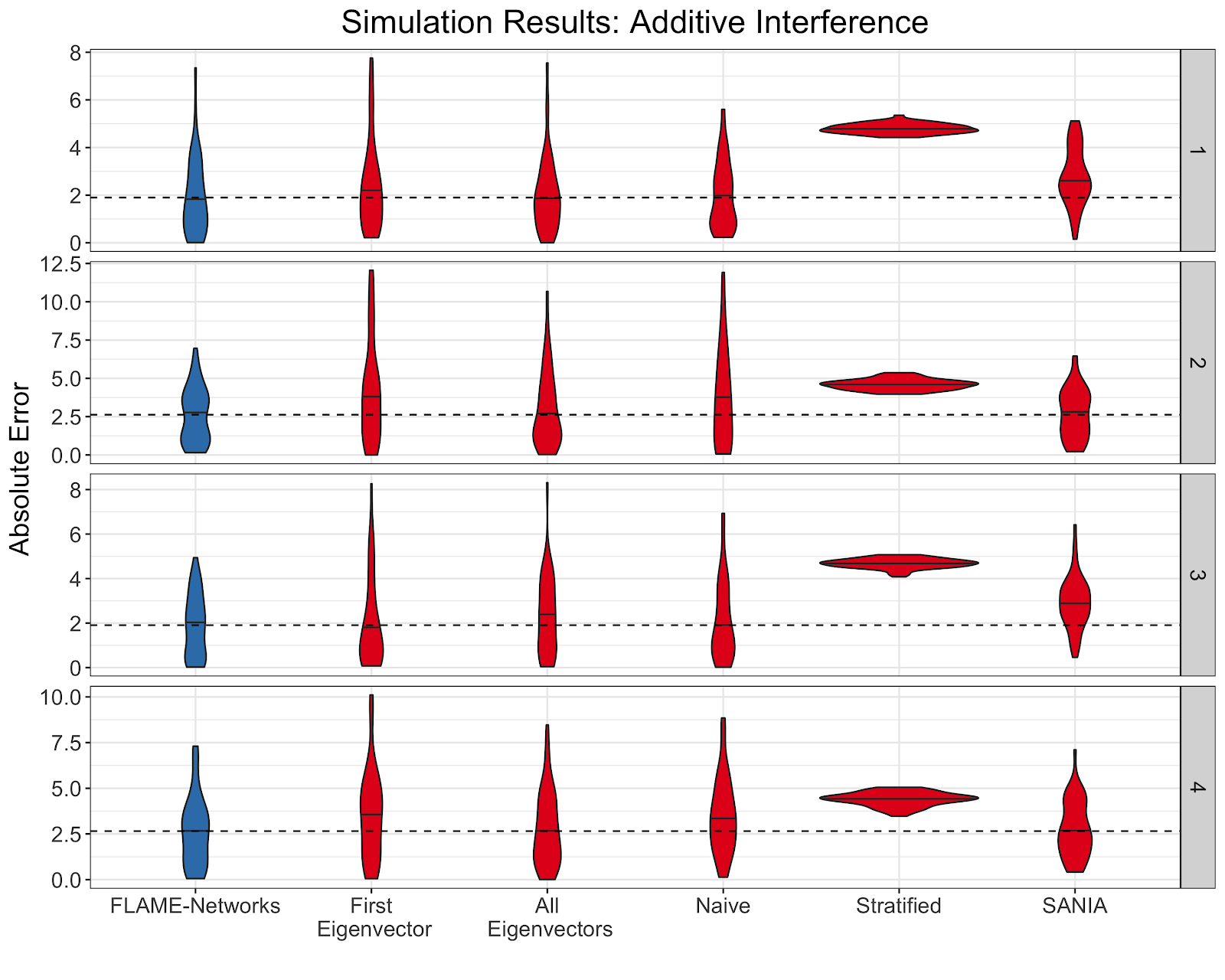}
    \caption{Results from experiments with additive interference on graphs in which treatment is randomly assigned within multiple clusters with few edges between them. Each violin plot represents the distribution over simulations of absolute estimation error over for each method. The panels are numbered according to the parameter setting the simulations were ran with. Violin plots are color-coded blue if the method had mean error either equal to or better than FLAME-Networks and red otherwise. The black line inside each violin is the median error. The dashed line is FLAME-Networks' mean error.}
    \label{Fig:cluster_rand}
\end{figure}

We also explored the performance of FLAME-Networks in settings in which treatment is randomized within multiple clusters, which have few connections between them. More specifically, we simulate a network according to a stochastic block model with 5 clusters. In each cluster, there are 10 units, 5 of which are treated. The probability of edges within clusters is 0.3 and between clusters is 0.05. This results in graphs with few edges between clusters. We then evaluate our method as previously described, simulating the outcome with additive interference and homoskedastic errors. The results in Figure \ref{Fig:cluster_rand} demonstrate that FLAME-Networks outperforms competing methods in this setting as well. 

\section{Additional Experiment: Real Network}\label{sec:realnetwork}
\begin{figure}[!htbp]
    \centering  
    \includegraphics[width=.99\linewidth]{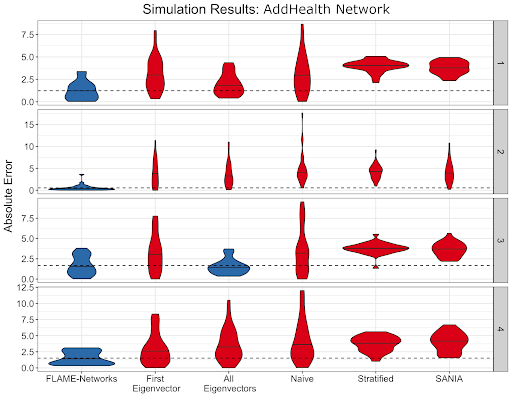}
    \caption{Results from experiments on a real, AddHealth network with additive interference.  Each violin plot represents the distribution over simulations of absolute estimation error over for each method. The panels are numbered according to the parameter setting the simulations were ran with. Violin plots are color-coded blue if the method had mean error either equal to or better than FLAME-Networks and red otherwise. The black line inside each violin is the median error. The dashed line is FLAME-Networks' mean error.}
    \label{Fig:addhealth}
\end{figure}

To ensure that FLAME-Networks also performs well on real networks, we consider an AddHealth network \citep{harris2009}. Specifically, we use the addhealthc3 dataset from the \pkg{amen} \textbf{\textsf{R}} package, with all edges treated as undirected. There are 32 nodes and on every simulation, 16 are randomly selected to be treated. Outcome and additive interference are simulated as previously described. Errors are homoskedastic. The results in Figure \ref{Fig:addhealth} demonstrate that FLAME-Networks still outperforms competing methods. 

\section{Additional Experiment: Heteroscedastic Errors}\label{sec:heteronetwork}

\begin{figure}[]
    \centering  
    \includegraphics[width=.99\linewidth]{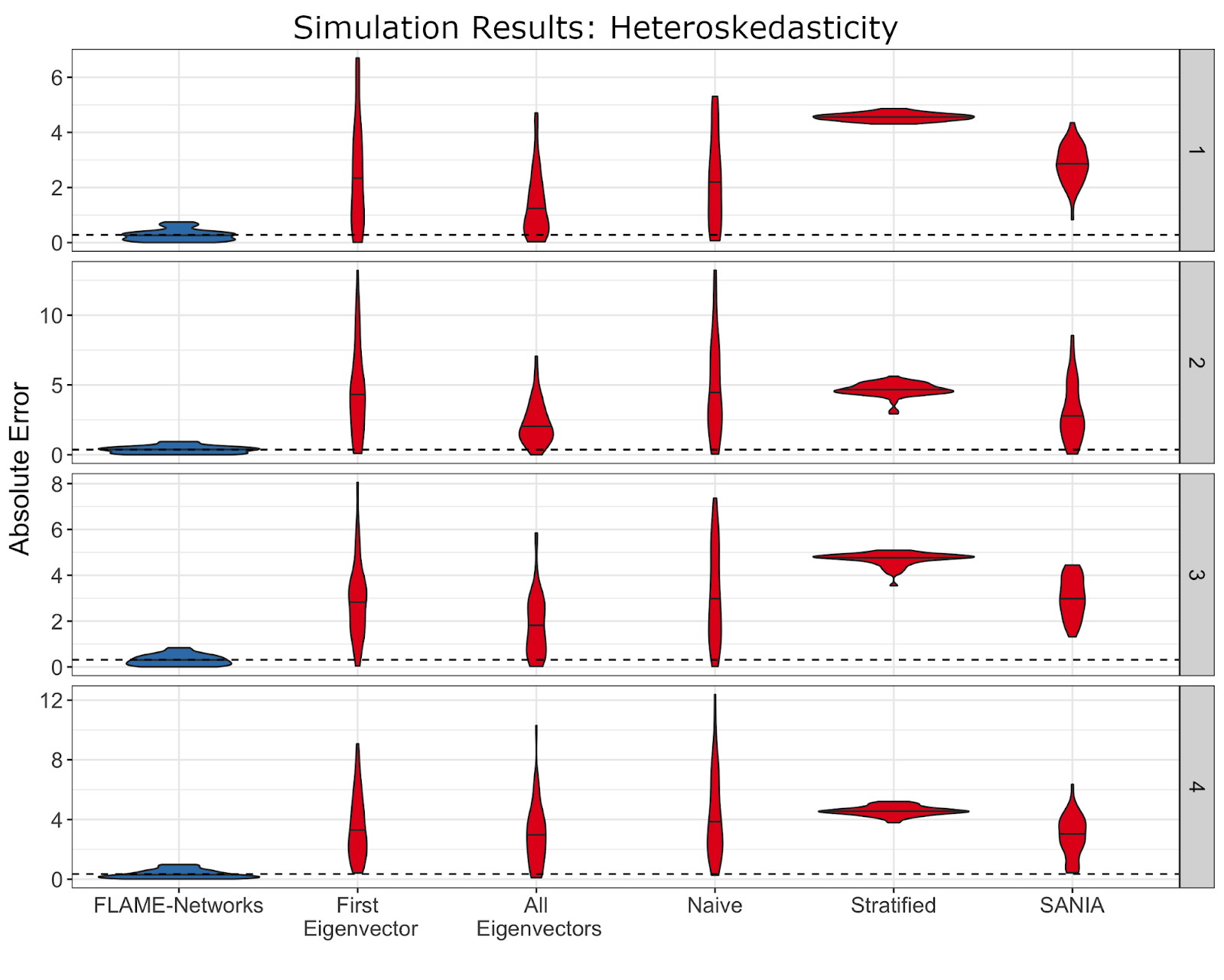}
    \caption{Results from experiments with an additive interference function involving heteroskedasticity in the baseline effects across units.  Each violin plot represents the distribution over simulations of absolute estimation error over for each method. The panels are numbered according to the parameter setting the simulations were ran with. Violin plots are color-coded blue if the method had mean error either equal to or better than FLAME-Networks and red otherwise. The black line inside each violin is the median error. The dashed line is FLAME-Networks' mean error.}
    \label{Fig:Heteroskedastic}
\end{figure}

We also explored the performance of FLAME-Networks in settings in which the variance of the outcomes is not constant. We simulate a single ER(50, 0.07) graph and randomly treat 25 units. We consider additive interference, as in the body of the text, and all other simulation parameters are the same, expect for that now, each unit $i$ has baseline outcome $\alpha_i \stackrel{ind}{\sim} N(0, v_i)$ with $v_i \stackrel{ind}{\sim} U(0, 1)$. We see in Figure \ref{Fig:Heteroskedastic} that FLAME-Networks outperforms competitors; the fact that it is nonparametric allows it to handle more flexible baseline outcomes and variances. 

\section{Additional Experiment: Matching on True Interference}\label{sec:trueinterference}
\begin{figure}[]
    \centering  
    \includegraphics[width=.99\linewidth]{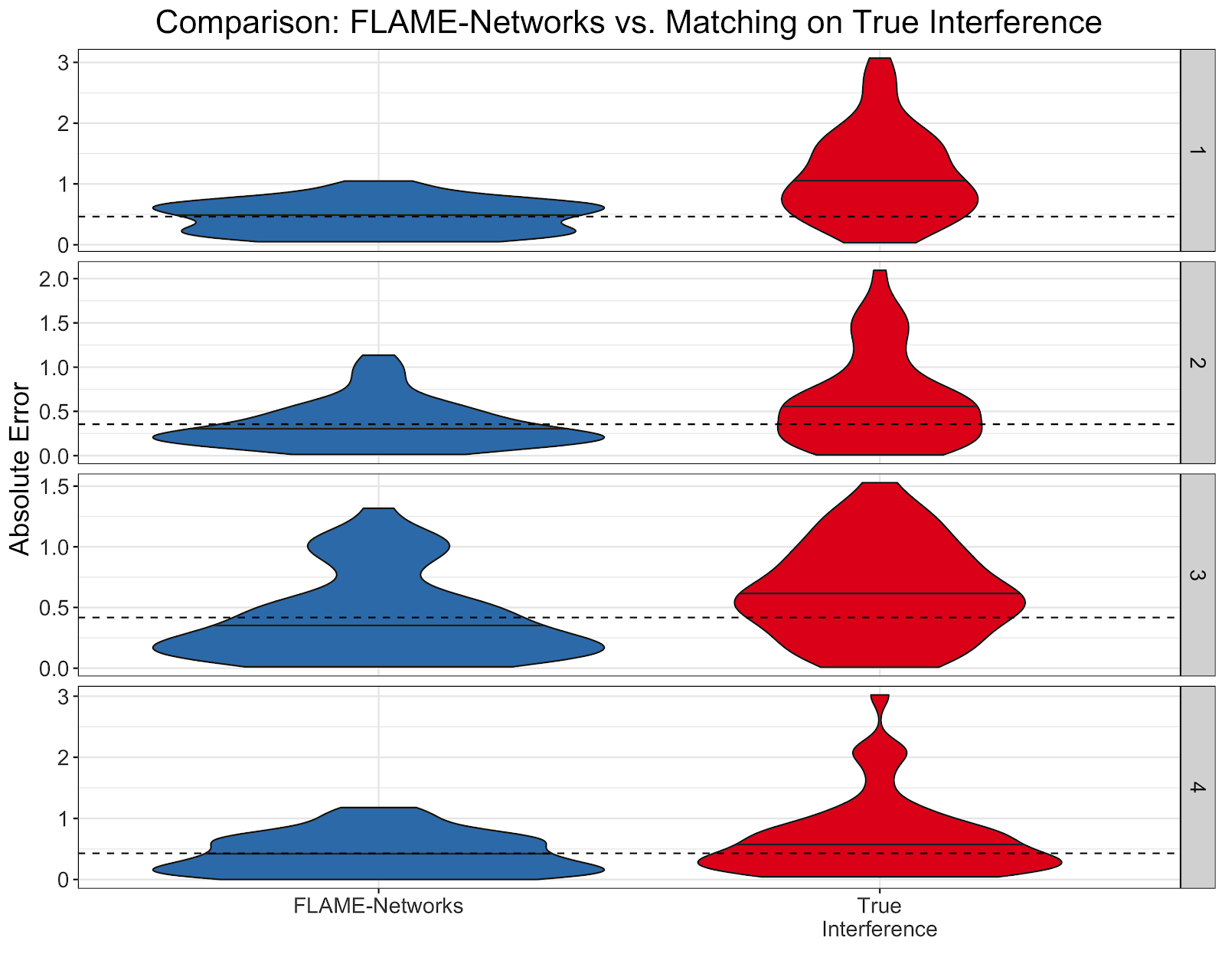}
    \caption{Results from experiments comparing FLAME-Networks to matching units on their true interference values. Each violin plot represents the distribution over simulations of absolute estimation error over for each method. The panels are numbered according to the parameter setting the simulations were ran with. Violin plots are color-coded blue if the method had mean error either equal to or better than FLAME-Networks and red otherwise. The black line inside each violin is the median error. The dashed line is FLAME-Networks' mean error.}
    \label{Fig:true_f}
\end{figure}

Here, we compare FLAME-Networks to approaches that match directly on units' interference values. FLAME-Networks already has the advantage of interpretably matching on neighborhood graphs that can be visualized and compared, as opposed to uninterpretable scalar values of an interference function. Additionally, to perform well in practice, one would typically need to use equally uninterpretable machine learning methods to estimate units' interference values well. But even comparing FLAME-Networks to an approach that matches on the \textit{true} (typically unknown) interference values, we see that our method does well in comparison, because it learns and matches on baseline effects as well as (approximate) interference values. Results using an ER(50, 0.07) graph with 25 units randomly treated, an additive interference function, and homoskedastic errors -- as previously described -- are shown in Figure \ref{Fig:true_f}.

\section{Additional Experiment: Covariate Weight}\label{sec:cov_weight}
In this section, we show that increasing the influence that covariates have on the outcome function harms neither FLAME-Networks nor the competing methods. As in the results shown in the main text, however, the performance of FLAME-Networks is still superior, given that it naturally handles covariate data. The experimental setup is the same as in Experiment 2 in the main text and results are shown in Tables \ref{Tab:more_cov_weight1} and \ref{Tab:more_cov_weight2}

\begin{table}[ht]
\centering

\begin{tabular}{lrrr}
  \hline
 Method & Median & 25th q & 75th q \\ 
  \hline
\textbf{FLAME-Networks} & 0.34 & 0.15 & 0.52 \\ 
\hline
  First
 Eigenvector & 0.41 & 0.24 & 0.49 \\ 
  All Eigenvectors & 0.36 & 0.32 & 0.74 \\ 
  Naive & 0.61 & 0.19 & 0.85 \\ 
  SANIA & 2.31 & 1.78 & 2.75 \\ 
  Stratified & 4.56 & 4.55 & 4.63 \\ 
  \hline
\end{tabular}
\caption{Additional results from the experimental setup of Experiment 2, but with $\beta = 20$. Median and 25th and 75th percentile of absolute error over 10 simulations.}
\label{Tab:more_cov_weight1}
\vspace{-2mm}
--\end{table}
\begin{table}[ht]
\centering

\begin{tabular}{lrrr}
  \hline
 Method & Median & 25th q & 75th q \\ 
  \hline
\textbf{FLAME-Networks} & 0.25 & 0.08 & 0.51 \\ 
\hline
  First
 Eigenvector & 0.52 & 0.32 & 0.85 \\ 
  All Eigenvectors & 0.53 & 0.29 & 0.83 \\ 
  Naive & 0.78 & 0.32 & 1.16 \\ 
  SANIA & 1.86 & 1.68 & 2.11 \\ 
  Stratified & 4.78 & 4.74 & 4.80 \\ 
  \hline
\end{tabular}
\caption{Additional results from the experimental setup of Experiment 2, but with $\beta = 25$. Median and 25th and 75th percentile of absolute error over 10 simulations.}
\label{Tab:more_cov_weight2}
\vspace{-2mm}
--\end{table}

\section{Match Quality}\label{sec:matchquality}
\begin{figure}[]
    \centering  
    \includegraphics[trim = {.65cm, 0, 0, 0}, clip, width=.99\linewidth]{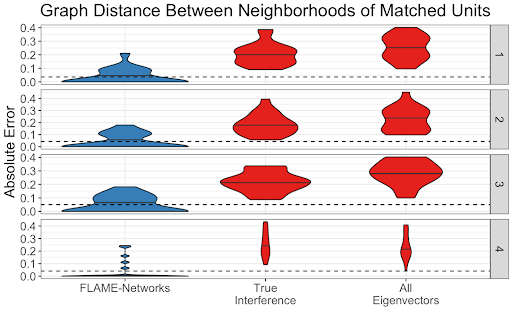}
    \caption{Results from experiments comparing the average distance between the neighborhood subgraphs of the units matched by different methods. Each violin plot represents the distribution over simulations of graph distance for each method. The panels are numbered according to the parameter setting the simulations were ran with. Violin plots are color-coded blue if the method had mean graph distance either equal to or better than FLAME-Networks and red otherwise. The black line inside each violin is the median graph distance. The dashed line is FLAME-Networks' mean graph distance.}
    \label{Fig:graph_dist}
\end{figure}

Here, we assess the quality of matches generated by FLAME-Networks versus matching on true interference, and the All Eigenvectors approach. To do so, for FLAME-Networks: for each (control) treated unit, we take the minimal Frobenius norm of the difference between that unit’s neighborhood adjacency matrix and that of all the (treated) control units in its matched group\footnote{The Frobenius norm of the difference of the adjacency matrices, up to reordering of the vertices.}, and average across all units. This gives an average graph distance for a single simulation. And to do so for the true interference matching and All Eigenvectors approaches: for every (control) treated unit, we take the closest (treated) control unit, find the graph distance (as above) between their neighborhood subgraphs, and average across units. This gives an average graph distance for a single simulation. Results from 50 simulations performed on an ER(50, 0.07) graph with additive interference and homoskedastic errors, as described previously, are shown in Figure \ref{Fig:graph_dist}, showing that FLAME-Networks produces more matches between units with similar neighborhood subgraphs than matching on the true interference or the All Eigenvectors method. 

\end{document}